\documentclass[a4paper,12pt]{article}
\usepackage{amssymb,amsmath}
\usepackage{mathrsfs}

\usepackage[titletoc]{appendix}

\allowdisplaybreaks \numberwithin{equation}{section}
\newtheorem{thm}{Theorem}[section]
\newtheorem{prp}[thm]{Proposition}
\newtheorem{lem}[thm]{Lemma}
\newtheorem{defn}[thm]{Definition}
\newenvironment{dfn}{\begin{defn} \rm }{\end{defn}}
\newtheorem{cor}[thm]{Corollary}
\newtheorem{remark}[thm]{Remark}

\newtheorem{example}{Example}[section]
\newenvironment{exa}{\begin{example} \rm }{ \end{example}}
\newenvironment{rmk}{\begin{remark} \rm }{\hfill $\Box$ \end{remark}}
\newenvironment{prf}{\noindent {\it Proof} \ }{\hfill $\Box$}
\newenvironment{prfof}[1]{\noindent {\it Proof of #1} \ }{\hfill $\Box$}

\newcommand\od{\mathrm{d}}
\newcommand\ad{\mathrm{ad}}

\newcommand{\nn}{\nonumber}

\newcommand{\diag}{\mathrm{diag}}
\newcommand\ep{\epsilon}
\newcommand\pd{\partial}
\newcommand{\ld}{\lambda} \newcommand{\Ld}{\Lambda}
\newcommand{\al}{\alpha}

\newcommand{\sg}{\sigma}
\newcommand{\om}{\omega} \newcommand{\Om}{\Omega}
\newcommand{\Gm}{\Gamma}
\newcommand{\Dt}{\Delta}
\newcommand{\dt}{\delta}
 \newcommand{\Ta}{\Theta}

\newcommand{\res}{\mathrm{res}}

\newcommand{\kn}{\mathrm{Ker}}
\newcommand{\im}{\mathrm{Im}}

\newcommand\C{\mathbb{C}}
\newcommand\R{\mathbb{R}}
\newcommand\Z{\mathbb{Z}}
\newcommand\Zop{\mathbb{Z^{\mathrm{odd}}_+}}

\newcommand\mL{\mathcal{L}}

\newcommand\sL{\mathscr{L}}

\newcommand\fg{\mathfrak{g}}

\newcommand\ra{\rangle}

\newcommand{\bt}{\mathbf{t}}  \newcommand{\bx}{\mathbf{x}}
\newcommand{\rs}{\mathrm{s}}

\begin{document}
\title{Tau Functions and Virasoro Symmetries for Drinfeld-Sokolov Hierarchies}
\author{Chao-Zhong Wu
\\
 {\small School of Mathematics and Computational Science, Sun Yat-sen University}
 \\
 {\small Guangzhou 510275, P.R. China.  }
}
\date{}
\maketitle

\begin{abstract}
For each Drinfeld-Sokolov integrable hierarchy associated to affine
Kac-Moody algebra, we obtain a uniform construction of tau function
by using tau-symmetric Hamiltonian densities, moreover, we represent
its Virasoro symmetries as linear/nonlinear actions on the tau
function. The relations between the tau function constructed in this
paper and those defined for particular cases of Drinfeld-Sokolov
hierarchies in the literature are clarified. We also show that,
whenever the affine Kac-Moody algebra is simply-laced or twisted,
the tau functions of the Drinfeld-Sokolov hierarchy coincide with
the solutions of the corresponding Kac-Wakimoto hierarchy
constructed from the principal vertex operator realization of the
affine algebra.

\end{abstract}

\tableofcontents

\section{Introduction}

For every affine Kac-Moody algebra $\fg$ with an arbitrary vertex of
its Dynkin diagram marked, Drinfeld and Sokolov \cite{DS}
constructed a hierarchy of integrable systems that generalize the
celebrated Korteweg-de Vries (KdV) equation. These integrable
hierarchies have very important applications in various areas of
mathematical physics like 2\,D topological field theory and
Gromov-Witten invariants \cite{DVV, Du, DZ, EYY, FSZ, FJR, Kon,
Witten}. For instance, the partition function of a topological
minimal model of ADE type is given by the logarithm of tau function
of the Drinfeld-Sokolov hierarchy associated to the corresponding
simply-laced affine Kac-Moody algebra. Such a tau function is
selected by the string equation, or equivalently, it must have
trivial evolution along the flow generated by the Virasoro symmetry
of level $-1$. In this paper we study tau functions of
Drinfeld-Sokolov hierarchies and their Virasoro symmetries.

In the literature there are several methods to define tau functions
for Drinfeld-Sokolov hierarchies (or their generalizations
\cite{dGHM}). One of them is based on certain integrable highest
weight representation of the affine Kac-Moody algebra $\fg$, see
\cite{HM, Im, Mi}. In this way, Drinfeld-Sokolov hierarchies are
closely related to the systems of Hirota bilinear equations
constructed by Date, Jimbo, Kashiwara and Miwa \cite{DKJM-KPBKP,
JM83}, and by Kac and Wakimoto \cite{Kac, KW}, meanwhile the tau
functions are identified with elements of the orbit space of the
highest weight vector acted by the affine Lie group. A shortcoming
of this method is that, it relies on the representation theory of
$\fg$, and usually involves some dressing operators, given
implicitly in a sense, of the hierarchies written in zero-curvature
form.

The second method is to define tau function via a family of
appropriate densities of Hamiltonians that are called to be
tau-symmetric in \cite{DZ}. Such Hamiltonian densities correspond to
some special two-point correlation functions whenever the logarithm
of the tau function gives a partition function in topological field
theory. This method, which does not depend on the representation
theory of Lie algebras, works well for the Drinfeld-Sokolov
hierarchies associated to affine algebras $A_n^{(1)}$. For example,
the $A_1^{(1)}$-type hierarchy is equivalent to the KdV hierarchy:
\begin{equation}\label{KdV}
\frac{\pd L}{\pd t_j}=[(L^{j/2})_+,L],\quad j\in\Zop,
\end{equation}
where $L=D^2+u$ with $D=\od/\od x$ and $u$ being a function of the
spatial variable $x$ and time variables $t_j$. The hierarchy
\eqref{KdV} has the following Hamiltonian representation
\begin{equation}\label{KdVham}
\frac{\pd u}{\pd t_j}=\{u(x), H_j\},
\end{equation}
in which the Poisson bracket reads
\[
\{u(x), u(y)\}=2 u(x)\dt'(x-y)+u'(x)\dt(x-y)+\frac1{2}\dt'''(x-y)
\]
and the Hamiltonian functionals are
\begin{equation*}\label{}
H_j=\int h_j(u; \pd_x u, \pd_x^2 u, \dots)\od x, \quad
h_j=\frac{2}{j}\res\,L^{j/2}.
\end{equation*}
The densities $h_j$ satisfy the tau-symmetry condition
\begin{equation}\label{GDht}
\frac{i}{2}\frac{\pd h_i}{\pd t_j}=\frac{j}{2}\frac{\pd h_j}{\pd
t_i}, \quad i, j\in\Zop,
\end{equation}
hence they define locally a tau function $\tau$ by
\begin{equation}\label{KdVtau}
\frac{\pd^2\log\tau}{\pd x\,\pd t_j}=\frac{j}{2}h_j, \quad j\in\Zop.
\end{equation}
In a similar way, we pushed forward the construction of tau function
to all $D_n^{(1)}$-hierarchies, see \cite{LWZ} or
Example~\ref{exa-Dn} below.

For an arbitrary Drinfeld-Sokolov hierarchy, however, it was unknown
how to choose such tau-symmetric Hamiltonian densities. The original
motivation of this paper is to resolve this problem universally,
rather than in a case-by-case way. Our first main result is \emph{a
uniform construction of tau function for each Drinfeld-Sokolov
hierarchy by}
\begin{equation}\label{DStau0}
\frac{\pd^2\log\tau}{\pd x\,\pd t_j}=-\frac{j(\Ld_j\mid
H)}{(\Ld_j\mid\Ld_{-j})}, \quad j\in E_+.
\end{equation}
Here $H$ is certain generating function of Hamiltonian densities of
the hierarchy, $t_j$ are the time variables corresponding to the
generators $\Ld_j$ of the principal Heisenberg algebra of $\fg$, for
which $E_+$ is the set of positive exponents and
$(\,\cdot\mid\cdot)$ is a nondegenerate invariant symmetric bilinear
form (see Section~\ref{sec-tau} for details). In particular, the tau
function given in \eqref{DStau0} is consistent with that defined in
the literature via Hamiltonian densities for each hierarchy of type
$A_n^{(1)}$ or $D_n^{(1)}$.

Besides the above two methods, the third important approach to
defined tau functions is to use certain line bundle on
infinite-dimensional Grassmannians. This approach dates back to Sato
and Segal-Wilson, who introduced tau functions for the $A_n^{(1)}$
case, see \cite{SW} and references therein. Based on Ben-Zvi and
Frenkel's geometric description \cite{BF} of (generalized)
Drinfeld-Sokolov hierarchies with smooth projective curves and Lie
groups, recently Safronov \cite{Sa} defined tau functions of these
hierarchies on a section of line bundle to the so-called
Drinfeld-Sokolov Grassmannians. He also pointed out that his tau
function coincides with $\tau$ in \eqref{DStau0} for the original
Drinfeld-Sokolov hierarchies. Following closely the approach in
\cite{SW}, we introduced in \cite{CW} tau functions of (generalized)
Drinfeld-Sokolov hierarchies starting from a reformulation of the
hierarchies with dressing operators, and showed that these tau
functions are equivalent to those given in \eqref{DStau0}. Here we
will not get into details of \cite{Sa, CW}, for they do not concern
the present paper.

\vskip 1ex

We continue to consider symmetries for Drinfeld-Sokolov hierarchies.
As the most simple case, the KdV hierarchy is known to possess a
family of so-called additional symmetries. These symmetries commute
with each flow in the hierarchy \eqref{KdV}, but do not commute
among themselves. Instead, they obey a Virasoro commutation
relation; that is why such additional symmetries are also called
Virasoro symmetries. More precisely, the Virasoro symmetries for the
KdV hierarchy are generated by the infinitesimal transformations
(see, for example, \cite{dV}) of tau function as
\begin{equation}\label{KdVvir}
\tau\mapsto \tilde{\tau}=\tau+\ep L_k\tau, \quad k\ge-1,
\end{equation}
where $\ep$ is a small parameter, and the generators $L_k$ reads
\begin{align}
&L_{-1}=\frac1{2}\sum_{j\in\Zop}(j+2)t_{j+2}\frac{\pd}{\pd
t_{j}}+\frac1{4}t_1^2,
\\
&L_{0}=\frac1{2}\sum_{j\in\Zop}j\,t_{j}\frac{\pd}{\pd
t_{j}}+\frac1{16}, \label{L0}
\\
&L_{k}=\frac1{4}\sum_{i=1}^k\frac{\pd^2}{\pd t_{2 i-1}\pd
t_{2k-2i+1}}+ \frac1{2}\sum_{j\in\Zop}j\,t_{j}\frac{\pd}{\pd
t_{j+2k}}, \quad k\ge1.
\end{align}
These operators satisfy
\begin{equation*}\label{}
[L_k, L_{l}]=(k-l)L_{k+l}, \quad k, l\ge-1.
\end{equation*}
In particular, the first two generators $L_{-1}$ and $L_0$
correspond to the Galilean and the scaling transformations
respectively. The string equation of the tau function is
\begin{equation}\label{}
\frac{\pd\tau}{\pd t_1}=L_{-1}\tau;
\end{equation}
it induces a series of constraints to $\tau$ that plays an important
role in topological field theory and matrix models \cite{AvM}.
Virasoro symmetries for Drinfeld-Sokolov hierarchy of type
$A_n^{(1)}$ or $D_n^{(1)}$ can be constructed by using
pseudo-differential operator skills, and they are written as linear
actions on tau function like \eqref{KdVvir}, see \cite{Wu-vir} for
the cases $D_n^{(1)}$. When the affine Kac-Moody algebra $\fg$ is
simply-laced, similar description of Virasoro symmetries for
Drinfeld-Sokolov hierarchies was given by Hollowood, Miramontes and
and S\'anchez Guill\'en \cite{HMSG}, who used the method of
representation theory of $\fg$.

So far as we know, there are no analogous characterizations of
Virasoro symmetries via tau function for arbitrary Drinfeld-Sokolov
hierarchies. The reason is probably the lack of an appropriate
definition of tau function of them before. As the tau function is
defined in \eqref{DStau0}, we will obtain another main result of
this paper.

\begin{thm}\label{thm-main}
Given an arbitrary affine Kac-Moody algebra $\fg$, the associated
Drinfeld-Sokolov hierarchy possesses Virasoro symmetries generated
by the following infinitesimal transformations of tau function:
\begin{equation}\label{tauvir0}
\tau\mapsto\tilde{\tau}=\tau+\ep (V_k\tau+\tau\,O_k), \quad k\ge-1,
\end{equation}
where $V_k$ are Virasoro operators independent of $\tau$, and $O_k$
are differential polynomials in second-order derivatives of\,
$\log\tau$ with respect to the time variables. Moreover, $O_k=0$ for
all $k\ge-1$ if $\fg$ is simply-laced or twisted, while
$O_{-1}=O_0=0$ if $\fg$ is of type B, C, F or G.
\end{thm}

This theorem provides a unified description of Virasoro symmetries
for all Drinfeld-Sokolov hierarchies. Firstly, if the affine
Kac-Moody algebra $\fg$ is simply-laced, the triviality of $O_k$
shows the \emph{linearization of Virasoro symmetries}. This agrees
with the previous results in \cite{dV, HMSG, DZ, Wu-vir}; for
example, $V_k=L_k$ whenever $\fg$ is of type $A_1^{(1)}$. As an
application of the definition of tau functions and the linearization
of Virasoro symmetries, Liu, Ruan and Zhang \cite{LRZ} proposed a
complete proof of the equivalence between the Drinfeld-Sokolov
hierarchies of simply-laced type and Dubrovin and Zhang's
topological hierarchies associated to semisimple Frobenius manifolds
for ADE-type simple singularities \cite{DZ} (see also \cite{DLZ,
LWZ, Wu-vir}).

Secondly, in case $\fg$ is of non-ADE type, we conjecture that the
functions $O_k$ with $k\ge1$ may not vanish; namely, the Virasoro
symmetries for Drinfeld-Sokolov hierarchies of non-ADE type are not
linearizable. This conjecture is partially verified (see
Example~\ref{exa-B2} and \cite{CW} for the $C_n^{(1)}$-hierarchies),
but is still open in general.

Although the Drinfeld-Sokolov hierarchies associated to twisted
affine Kac-Moody algebras contain important examples such like the
Sawada-Kotera equation \cite{SK} (belonging to the
$A_2^{(2)}$-hierarchy), they seem not have attracted much attention
in a sense. In fact, when the affine Kac-Moody algebra $\fg$ is
twisted, the Drinfeld-Sokolov hierarchy is Hamiltonian \cite{DS},
and probably has only one (local) Hamiltonian structure
\cite{LWZ-ham, SK} such that it is not involved in the framework of
topological hierarchies in \cite{DZ}. What is surprising, we now
show that this hierarchy has a tau function defined by Hamiltonian
densities, as well as linearized Virasoro symmetries acting on the
tau function. For such kind of hierarchies, it is unknown whether
there is any illustration in topology that is analogous with the
case of hierarchies of simply-laced type.

Our proof of linearization of Virasoro symmetries is based on the
representation theory of simply-laced or twisted affine Kac-Moody
algebras. This naturally leads us to study the relation between
Drinfeld-Sokolov hierarchies and Kac-Wakimoto hierarchies of
bilinear equations. Recall that Drinfeld-Sokolov hierarchies for
simply-laced affine algebras were shown related to the corresponding
Kac-Wakimoto hierarchies by Hollowood and Miramontes \cite{HM}, see
their formula \eqref{hmtau} below. Inspired by their work, we obtain
a byproduct of the present paper, that is, \emph{if $\fg$ is a
simply-laced or twisted affine Kac-Moody algebra, then tau functions
given in \eqref{DStau0} of the Drinfeld-Sokolov hierarchy coincide
with solutions of the Kac-Wakimoto hierarchy constructed from the
principal vertex operator realization of $\fg$} (see
Theorem~\ref{thm-KW} below).

\vskip 1ex

To achieve the above results, we organize the contents of this paper
as follows.

In the forthcoming section, we will recall Drinfeld and Sokolov's
original construction of integrable hierarchy on a ``loop algebra'',
that is, the affine Kac-Moody algebra $\fg$ modulo the subspace
spanned by the central and the scaling elements $c$ and $d$. The
hierarchy is composed of Hamiltonian equations, with Hamiltonian
densities being determined up to addition of total derivatives with
respect to the spacial variable.

In Section~3, we reformulate the definition of Drinfeld-Sokolov
hierarchies on the derived algebra $\fg'$ of $\fg$. The nontrivial
central part helps us to fix the freedom of Hamiltonian densities to
fulfill the tau-symmetry condition, hence a tau function can be
defined by \eqref{DStau0}.

In Section~4, we will review the Kac-Moody-Virasoro algebra
consisting of $\fg$ and a family of derivations on it. The
construction of Virasoro symmetries in \cite{HMSG} will be revised
and extended to all Drinfeld-Sokolov hierarchies. Moreover, the
Virasoro symmetries will be represented via tau function in a
unified form \eqref{tauvir0}, and they are shown to be linearized in
case $\fg$ is simply-laced or twisted, which proves
Theorem~\ref{thm-main}.

Section~5 is a collection of examples. The first three examples show
the consistence between the tau function in \eqref{DStau0} and those
defined in the literature for Drinfeld-Sokolov hierarchies of types
$A_n^{(1)}$ and $D_n^{(1)}$. The other examples illustrate how to
compute $O_k$ in \eqref{tauvir0} that give obstacles when
linearizing the Virasoro symmetries. The Virasoro constraints to tau
function will also be derived.

The last section is devoted to the conclusion and some discussions.
We try to divide Drinfeld-Sokolov hierarchies into three classes
according to their Hamiltonian structures and Virasoro symmetries,
and discuss possible applications.

\vskip 1ex

In order to make this paper more complete, in the appendix we will
consider tau functions of integrable hierarchies of modified KdV
type from Drinfeld and Sokolov's construction \cite{DS}. To avoid
lengthy expressions, we call such hierarchies the \emph{modified
Drinfeld-Sokolov hierarchies}, which are related to the
Drinfeld-Sokolov hierarchies (of KdV type) by certain gauge
transformations. Note that for untwisted affine Lie algebras, such
modified hierarchies were also constructed by Kupershmidt and Wilson
\cite{KWi, Wi}. Another equivalent version of these hierarchies was
given by Feigin and Frenkel \cite{FF}; accordingly Enriquez and
Frenkel \cite{EF} introduced tau functions of them with the help of
tau-symmetric Hamiltonian densities.

It will be seen that the right hand side of \eqref{DStau0} is
invariant with respect to gauge transformations, hence $\tau$ also
serves as a tau function of the corresponding modified
Drinfeld-Sokolov hierarchy. This tau function will be shown
different from the one defined by Enriquez and Frenkel. As a matter
of fact, in the modified case, these two tau functions coincide with
two special cases of the tau function obtained by Miramontes
\cite{Mi} based on the representation theory of $\fg$, see
Proposition~\ref{thm-taum} below.

\section{Definition of Drinfeld-Sokolov hierarchies}
\label{sec-DS}

Given an affine Kac-Moody algebra, Drinfeld and Sokolov's
hierarchies associated to different marked vertices of the Dynkin
diagram are related by Miura-type transformations. For convenience,
in this paper we only consider the case that the vertex is chosen to
be the zeroth one, which is the special vertex added to the Dynkin
diagram of the corresponding simple Lie algebra.

\subsection{Properties of affine Kac-Moody algebras}

Let $A=(a_{ij})_{0\le i, j\le n}$ be a generalized Cartan matrix of
affine type, and $\fg(A)$ be the corresponding Kac-Moody algebra.
Recall that $\fg(A)$ is generated by a set of Weyl generators
\[
\{e_i, f_i, \al^\vee_i \mid i=0, 1, 2, \ldots, n\}
\]
and a scaling element $d$. One has the decomposition
$\fg(A)=\fg'(A)\oplus\C d$, with $\fg'(A)$ being the derived
algebra. The center of $\fg(A)$ (or of $\fg'(A)$) is spanned by the
canonical central element, say, $c$, which satisfies
\begin{equation}
c=\sum_{i=0}^n k_i^\vee \al^\vee_i.
\end{equation}
Here $k_i^\vee$ are the dual Kac labels of $\fg(A)$, i.e., the
lowest positive integers that solve the linear equation
$\sum_{i=0}^n k_i^\vee a_{ij}=0$. For the sake of simplifying
notations, we will write $\fg=\fg(A)$ and $\fg'=\fg'(A)$ below.

An arbitrary integer vector $\rs=(s_0, s_1, \dots, s_n)\in\Z^{n+1}$,
with $s_i\ge0$ but not all equal to $0$, induces a gradation on
$\fg'$ by setting
\begin{equation}\label{sgr}
\deg e_i=s_i, \quad \deg f_i=-s_i, \quad \deg \al_i^\vee=0.
\end{equation}
The following two gradations are of particular importance \cite{DS,
Kac}:
\begin{itemize}
\item[(i)] the homogeneous/standard gradation
\begin{equation}\label{hmgr}
\fg'=\bigoplus_{j\in\Z}\fg'_j ~\hbox{  induced by  }~
\rs^0=(1,0,\dots,0);
\end{equation}
\item[(ii)] the
principal/canonical gradation
\begin{equation}\label{prgr}
\fg'=\bigoplus_{j\in\Z}\fg'^j~\hbox{  induced by }~
\rs^1=(1,1,\dots,1).
\end{equation}
\end{itemize}
Conventions like $\fg'_{\ge0}=\sum_{i\ge0}\fg'_i$ and
$\fg'^{<0}=\sum_{i<0}\fg'^i$ will be used below.

Let $E$ be the set of exponents of $\fg'$. In $\fg'$ there is a
so-called principal Heisenberg subalgebra $\mathfrak{s}$, which has
a basis $\{c, \Ld_j\in\fg'^j\mid j \in E\}$ such that
\begin{equation}\label{Ldij}
[\Ld_i, \Ld_j]=\dt_{i,-j}\,i\cdot c.
\end{equation}
In particular, $1$ is always an exponent, and $\Ld_1=\nu\Ld$ for
some nonzero constant $\nu$, where $\Ld=\sum_{i=0}^n  e_i$. The
element $\Ld$ induces the following decomposition of subspaces:
\begin{equation}\label{dec2}
\fg'=\mathfrak{s}+\im\,\ad_{\Ld}, \quad
\mathfrak{s}\cap\im\,\ad_{\Ld}=\C\,c.
\end{equation}
This property is crucial in the construction of Drinfeld-Sokolov
hierarchies.

\subsection{Drinfeld-Sokolov hierarchies}

Drinfeld and Sokolov's original construction works on the centerless
affine Lie algebra $\bar{\fg}=\fg'/\C\,c$. This algebra is graded in
the same way as for $\fg'$. Similarly, the homogeneous and principal
gradations are written respectively as
\[
\bar\fg=\bigoplus_{j\in\Z}\bar\fg_j, \quad
\bar\fg=\bigoplus_{j\in\Z}\bar\fg^j.
\]
Clearly $\bar\fg_0=\mathring{\fg}$, which denotes the simple Lie
algebra for the Cartan matrix $\mathring{A}=(a_{i j})_{1\le i, j\le
n}$ of finite type.

According to the decomposition \eqref{dec2}, the centralizer of
$\Ld=\sum_{i=0}^n e_i$ in $\bar\fg$ is the principal Heisenberg
subalgebra $\bar{\mathfrak{s}}=\mathfrak{s}/\C\,c$ of trivial
center. This subalgebra contains a basis $\{\Ld_j\in\bar\fg^j\mid j
\in E\}$ chosen as before.

We use $C^\infty(\R,W)$ to denote the set of smooth functions from
$\R$ to some linear space $W$ (the space $\R$ is not essential; it
can be replaced by other $1$-dimensional spaces such like the unit
circle $S^1$). Consider operators of the form
\begin{equation}\label{msL}
\sL= D+\Ld + q, \quad q\in C^\infty(\R,
\mathring{\fg}\cap\bar\fg^{\le0}),
\end{equation}
where $D={\od}/{\od x}$ with $x$ being the coordinate of $\R$.
Observe that $q$ is a smooth function taking value in the Borel
subalgebra of the simple Lie algebra $\mathring{\fg}$ generated by
$\al_i^\vee$ and $f_i$ with $i=1,\dots,n$. For operators of the form
\eqref{msL}, there are gauge transformations defined by
\begin{equation}\label{gauge}
\sL\mapsto e^{\ad_N}\sL,  \quad N\in C^\infty(\R,
\mathring{\fg}\cap\bar\fg^{<0}).
\end{equation}
In other words, this is an action of the Lie group of the nilpotent
subalgebra of $\mathring{\fg}$, which has an $n$-dimensional orbit
space.

The following proposition plays a fundamental role in the
construction of Drinfeld-Sokolov hierarchies.
\begin{prp}[\cite{DS}]\label{thm-dr0}
There exists a function~$U\in C^\infty(\R, \bar\fg^{< 0})$ such that
the operator $\bar{\sL}=e^{-\ad_U}\sL$ has the form
\begin{equation}\label{}
\bar{\sL}= D+\Ld+H, \quad H\in C^\infty(\R,
\bar{\mathfrak{s}}\cap\bar\fg^{<0}).
\end{equation}
Suppose $\tilde{U}$ also satisfies the above condition, then
$e^{-\ad_{\tilde{U}} } e^{\ad_U}=e^{\ad_S}$ with some $S\in
C^\infty(\R, \bar{\mathfrak{s}}\cap\bar\fg^{<0})$. Moreover, for
different choices of $U$, the function $H$ differs by adding the
total derivative of a differential polynomial in (components of)
$q$.
\end{prp}

According to the above proposition, one can choose a function $U$
and introduce a map
\begin{align}\label{}
\varphi:C^\infty(\R, \bar\fg)&\to C^\infty(\R, \bar\fg), \nn\\
X&\mapsto e^{\ad_U} X. \label{phi0}
\end{align}

\begin{defn}[\cite{DS}]\label{def-DS}  \rm
The Drinfeld-Sokolov hierarchy associated to $\bar\fg$ and the
zeroth vertex of its Dynkin diagram is the following family of
partial differential equations
\begin{equation}\label{Lt2}
\frac{\pd \sL}{\pd t_j}=[-\varphi(\Ld_j)_{\ge0}, \sL],  \quad j\in
E_+
\end{equation}
restricted to some equivalence class of $\sL$ with respect to the
gauge transformations \eqref{gauge}. Here the subscript ``$\ge0$''
means the projection to $\bar\fg_{\ge0}$, and $E_+$ is the set of
positive exponents.
\end{defn}

Drinfeld-Sokolov hierarchies restricted to different gauge slices of
$\sL$ are equivalent up to a gauge transformation of the form
\eqref{gauge}. In particular, if the gauge slice is given by $q$
taking value in the Cartan subalgebra of $\mathring{\fg}$, then we
call the corresponding hierarchy the \emph{modified Drinfeld-Sokolov
hierarchy}. This name is from the fact that in the modified case the
first nontrivial equation in the hierarchy for $A_1^{(1)}$ is the
modified KdV equation, which is related to the KdV equation by the
Miura transformation, see Example~\ref{exa-EF} below. One can refer
to \cite{KWi, Wi, FF, EF} for equivalent versions of such modified
hierarchies associated to untwisted affine Kac-Moody algebras.

Consider formal functionals of the form
\[
\mathscr{F}=\int f\left(q; \pd_x q, \pd_x^2 q, \dots\right)\,\od x
\]
that are invariant under the gauge transformations \eqref{gauge}.
Note that such a functional is not really an integral but formally
defined up to addition of total derivatives to the density $f$. The
gradient of a functional $\mathscr{F}$ with respect to $q$ is
defined to be $\mathrm{grad}_q\mathscr{F}\in
C^\infty(\R,\mathring{\fg}\cap\bar\fg^{\ge0})$ such that
\begin{equation*}
\left.\frac{\od }{\od \ep}\right|_{\ep=0}\mathscr{F}(q+\ep\,
\tilde{q})=\int(\mathrm{grad}_q\mathscr{F}\mid \tilde{q})\od x
\end{equation*}
for arbitrary $\tilde{q}\in
C^\infty(\R,\mathring{\fg}\cap\bar\fg^{\le0})$, where
$(\,\cdot\mid\cdot)$ is a nondegenerate invariant symmetric bilinear
form on $\bar\fg$.

There is a Poisson bracket between the gauge invariant functionals:
\begin{align} \label{dspoi2}
\{\mathscr{F},\mathscr{G}\}(q)
=\int\left(\mathrm{grad}_q\mathscr{F}\mid\Big[\mathrm{grad}_q\mathscr{G},D+\sum_{i=1}^{n}
e_i +q\Big]\right)\od x.
\end{align}
\begin{thm} [\cite{DS}] \label{thm-dsham}
The Drinfeld-Sokolov hierarchy \eqref{Lt2} can be written in a
Hamiltonian form as
\begin{equation}\label{DSbh2}
\frac{\pd\mathscr{F}}{\pd t_j}=\{\mathscr{F},\mathscr{H}_{j}\},
\quad j\in E_+,
\end{equation}
where the Hamiltonians are
\begin{equation}\label{sHj}
\mathscr{H}_j=\int (-\Ld_j\mid H)\,\od x
\end{equation}
with $H$ given in Proposition~\ref{thm-dr0}.
\end{thm}

\begin{rmk}
When the affine Lie algebra $\bar\fg$ is untwisted, the
Drinfeld-Sokolov hierarchy possesses another Hamiltonian structure
that is compatible with \eqref{DSbh2}. In other words, it is a
hierarchy of bi-Hamiltonian systems. The bi-Hamiltonian structure
was shown to be characterized by a semisimple Frobenius manifold
\cite{Du} on the orbit space of the corresponding Weyl group
together with a class of constant central invariants \cite{DLZ}.
\end{rmk}

In \cite{DS}, Drinfeld and Sokolov did not considered tau functions
of their hierarchies. Instead, they proposed a scheme to represent
their hierarchies into Lax equations of scalar pseudo-differential
operators. For example, the hierarchy \eqref{Lt2} for the affine
Kac-Moody algebra of type $A_1^{(1)}$ can be written equivalently to
the KdV hierarchy \eqref{KdV}. In summary, Drinfeld and Sokolov
obtained the Lax representations for the hierarchies \eqref{Lt2}
associated to the affine Kac-Moody algebras of types $A_n^{(1)}$,
$B_n^{(1)}$, $C_n^{(1)}$, $A_{2n}^{(2)}$, $A_{2n-1}^{(2)}$ and
$D_{n+1}^{(2)}$; for the hierarchy of type $D_{n}^{(1)}$, a Lax
representation was partially given in \cite{DS}, and completed by us
in \cite{LWZ} with the help of certain extended pseudo-differential
operators. Generally speaking, based on the Lax representations, tau
function of such hierarchies can be introduced via Hamiltonian
densities chosen similarly as for the KdV hierarchy, see, for
example, \cite{LWZ, CW} and Examples~\ref{exa-An}--\ref{exa-Dn}
below. However, this is a case-by-case method and may fail to work
in general. In the next section we will propose a way to construct
tau functions for all Drinfeld-Sokolov hierarchies.

\section{Tau function of Drinfeld-Sokolov hierarchies}
\label{sec-tau}

Given an operator $\sL$ in \eqref{msL}, the Hamiltonian densities in
\eqref{sHj} are defined up to addition of the total derivative of
differential polynomials in $q$, which depend on the function $U$ in
Proposition~\ref{thm-dr0}. Our idea is to fix the function $U$
appropriately such that the Hamiltonian densities are tau-symmetric,
hence a tau function can be defined. To this end, we will first
reformulate the Drinfeld-Sokolov construction on the derived algebra
$\fg'=\bar\fg\oplus\C\,c$, as inspired by \cite{HM}.

\subsection{Reformulation of Drinfeld-Sokolov hierarchies}

Recall that the operator in \eqref{msL} is just
\begin{equation}\label{sLgp}
\sL= D+\Ld + q, \quad q\in C^\infty(\R,
\mathring{\fg}\cap\fg'^{\le0}).
\end{equation}
In comparison with Proposition~\ref{thm-dr0}, we have the following
\begin{prp}\label{thm-dr}
Given an operator $\sL$ as \eqref{sLgp}, there is a unique function
$U\in C^\infty(\R, \fg'^{< 0})$ satisfying the following two
conditions
\begin{itemize}
\item[(i)] The operator $\bar{\sL}=e^{-\ad_U}\sL$ has the form
\begin{equation}\label{L0g}
\bar{\sL}= D+\Ld+H, \quad H\in C^\infty(\R,
\mathfrak{s}\cap\fg'^{<0});
\end{equation}
\item[(ii)] For every positive exponent $j\in E_+$, the central part of $e^{\ad_U}\Ld_j$
vanishes, namely
\begin{equation}
\label{ULdc} \big(e^{\ad_U}\Ld_j\big)_c=0.
\end{equation}
Here the subscript ``$c$'' means to take the coefficient of the
center $c$ with respect to the decomposition
$\C\al_1^\vee\oplus\cdots\oplus\C\al_n^\vee\oplus\C\,c$ of the
Cartan subalgebra of $\fg'$.
\end{itemize}
Moreover, both $U$ and $H$ are differential polynomials in $q$.
\end{prp}
\begin{prf}
One writes $e^{\ad_U}\bar{\sL}=\sL$ to
\begin{equation}
e^{\ad_{\sum_{k\le-1}}U_k}\left(D+\Ld+\sum_{k\le-1}H_k\right)=D+\Ld
+ \sum_{k\le0}q_k,
\end{equation}
where $q_k$, $U_k$, $H_k$ take value in $\fg'^k$. By comparing the
homogeneous terms we have
\begin{align}\label{Um1}
[U_{-1}, \Ld]&=q_0, \\
H_{k+1}+[U_{k},\Ld]&=\ast, \quad k=-2,-3,-4,\dots. \label{HUk}
\end{align}
Here for every $k$ the right hand side of \eqref{HUk} depends on
$q_k$, $H_i~ (i>k+1)$ and $U_i~ (i>k)$.

First of all, equation \ref{Um1} has a unique solution $U_{-1}$, for
which equally \eqref{ULdc} with $j=1$ is valid automatically. When
$k<-1$, by virtue of the decomposition \eqref{dec2}, the functions
 $H_{k+1}$ and $U_{k}$ can be solved recursively from \eqref{HUk}. In
more details, suppose $H_i~ (i>k+1)$ and $U_i~ (i>k)$ are given,
then $H_{k+1}$ is determined uniquely due to the decomposition
\eqref{dec2}. In finding $U_{k}$ there are two cases: first, the
function $U_{k}$ is unique whenever $k\not\in E$; second, if $k\in
E$, then $U_k$ is determined up to addition of a multiple of
$\Ld_{k}$. But the freedom in the latter case is fixed precisely by
the condition \eqref{ULdc} with $j=-k$. Therefore the proposition is
proved.
\end{prf}

one can refer to Example~\ref{exa-A1} below for an illustration of
the role played by the condition \eqref{ULdc} in calculating the
functions $U$ and $H$.

\begin{rmk}
In \cite{HM} Hollowood and Miramontes fix the function $U$ in a
different way. They let $U$ take value in
$\mathfrak{s}^\bot\cap\fg'^{<0}$, where $\mathfrak{s}^\bot$ is the
orthogonal complement of $\mathfrak{s}$ with respect to the standard
bilinear form on $\fg'$, see Proposition~2.1 in \cite{HM}. For such
a $U$, the corresponding Hamiltonian densities in \eqref{sHj} are
not what we look for.
\end{rmk}

\begin{lem} \label{thm-UN}
The function $H$ in Proposition~\ref{thm-dr} is invariant with
respect to the gauge transformations \eqref{gauge}.
\end{lem}
\begin{prf}
Assuming $\sL$ to be of the form \eqref{sLgp}, we have $U$ and $H$
determined by Proposition~\ref{thm-dr}. For any
$\tilde{\sL}=e^{\ad_N}\sL$ with $N\in
C^\infty(\R,\mathring{\fg}\cap\fg'^{<0})$, one has
\begin{equation*}
\tilde{\sL}=e^{\ad_{\tilde{U}}}(D+\Ld+H), \quad
e^{\ad_{\tilde{U}}}=e^{\ad_N}e^{\ad_U},
\end{equation*}
where $\tilde{U}\in C^\infty(\R,\fg'^{<0})$. We need to show that
$\tilde{U}$ also satisfies the condition \eqref{ULdc}. In fact, note
$[N, X]_c=0$ for any $X\in \fg'$, hence
\begin{equation*}\label{}
\big(e^{\ad_{\tilde{U}}}\Ld_j\big)_c=e^{\ad_N}\big(e^{\ad_U}\Ld_j\big)_c=0,\quad
j\in E_+.
\end{equation*}
The lemma is proved.
\end{prf}

Given an operator $\sL$ in \eqref{sLgp}, henceforth we fix the
functions $U$ and $H$ as in Proposition~\ref{thm-dr}. Similar to
\eqref{phi0} we now have
\begin{align}\label{}
\varphi:C^\infty(\R, \fg')&\to C^\infty(\R, \fg'), \nn\\
X&\mapsto e^{\ad_U} X. \label{phi}
\end{align}
By virtue of the condition \eqref{ULdc}, the evolutionary equations
\eqref{Lt2} are still well defined with $\bar\fg$ replaced by $\fg'$
and simultaneously the subscript ``$\ge0$'' becomes the projection
$\fg'\to\fg'_{\ge0}$. These equations compose the Drinfeld-Sokolov
hierarchy when restricted to the gauge equivalence class of $\sL$
with respect to the transformations \eqref{gauge}.

\subsection{Definition of tau function}

In order to define tau function of the Drinfeld-Sokolov hierarchy
\eqref{Lt2}, we introduce a nonlocal function
\begin{equation}\label{}
\Om=-\int^x H(q; \pd_x q, \pd_x^2 q, \dots)\,\od x.
\end{equation}
Namely, $\Om$ satisfies
\begin{equation}\label{Omx}
\frac{\pd\Om}{\pd x}=-H,
\end{equation}
and takes the form
\begin{equation}\label{OmH}
\Om=\sum_{j\in E_+}\frac{\om_j}{j}\Ld_{-j},
\end{equation}
where $\om_j$ are scalar functions determined up to addition of
constants.

\begin{lem}\label{thm-omij}
For the scalar functions $\om_j$ given above, all derivatives
$\pd\om_j/\pd t_i$ are differential polynomials in $q$, and they
satisfy
\begin{equation}\label{omtij}
\frac{\pd\om_j}{\pd t_i}=\frac{\pd\om_i}{\pd t_j}, \quad i,j\in E_+.
\end{equation}
\end{lem}
\begin{prf}
Recalling $\sL=e^{\ad_U}(D+\Ld+H)$, the following identity can be
verified straightforwardly (see, for example, Lemma~A.1 in
\cite{TT}):
\begin{equation}\label{sLtj}
 \frac{\pd\sL}{\pd t_j}=e^{\ad_U}\frac{\pd H}{\pd
t_j}+\left[ \nabla_{t_j, U} U, \sL\right]
\end{equation}
where
\[
\nabla_{t_j, U} U=\sum_{m\ge0}\frac1{(m+1)!}(\ad_U)^m\frac{\pd
U}{\pd t_j}.
\]
This together with \eqref{Lt2} leads to
\[
e^{\ad_U}\frac{\pd H}{\pd t_j}+[\nabla_{t_j, U}
U-\varphi(\Ld_j)_{<0}, \sL] = [-\varphi(\Ld_j),\sL],
\]
namely,
\begin{equation}\label{HU}
 \frac{\pd H}{\pd t_j}+[e^{-\ad_U}\left(\nabla_{t_j, U}
U-\varphi(\Ld_j)_{<0}\right), D+\Ld+H] = \frac{\pd\om_j}{\pd x}\cdot
c.
\end{equation}
Here on the right hand side we have used the condition \eqref{ULdc}.

According to the decomposition \eqref{dec2}, we write
\[
e^{-\ad_U}\left(\nabla_{t_j, U}
U-\varphi(\Ld_j)_{<0}\right)=G+\tilde{G}
\]
with $G$ and $\tilde{G}$ taking value in $\mathfrak{s}\cap\fg'^{<0}$
and in $\im\,\ad_\Ld\cap\fg'^{<0}$ respectively. Equation \eqref{HU}
splits into three parts:
\begin{align}
\label{HG}
&\frac{\pd H}{\pd t_j}-\frac{\pd G}{\pd x}=0, \\
&[G,\Ld]_c=\frac{\pd\om_j}{\pd x}, \label{GLd} \\
&-\frac{\pd \tilde{G}}{\pd x}+[\tilde{G},\Ld+H]=0.
\end{align}

Firstly, equation \eqref{HG} together with \eqref{Omx} implies
\begin{equation}\label{}
\frac{\pd\Om}{\pd t_j}=G, \quad j\in E_+.
\end{equation}
Hence
\begin{equation}\label{omitj}
\frac{\pd\om_i}{\pd t_j}=\left[\Ld_i, \frac{\pd\Om}{\pd
t_j}\right]_c=[\Ld_i,G]_c
\end{equation}
are differential polynomials in $q$.

In particular, taking $i=1$ in \eqref{omitj},  from $\Ld_1=\nu\Ld$
and \eqref{GLd}  it follows that
\begin{equation}\label{om1tj}
\frac{\pd\om_1}{\pd t_j}=\nu\frac{\pd\om_j}{\pd x}, \quad j\in E_+.
\end{equation}
Hence for $i,j\in E_+$, we obtain
\[
\frac{\pd^2\om_j}{\pd t_i\pd x}=\frac{\pd^2\om_i}{\pd t_j\pd x}.
\]
Both sides of the equality are total derivatives of differential
polynomials in $q$ with respect to $x$, hence it leads to
\eqref{omtij} by integration. The lemma is proved.
\end{prf}

Given a solution of the hierarchy \eqref{Lt2}, there locally exists
a smooth function $\tau$ of $\bt=(t_j)_{j\in E_+}$ such that
  \begin{equation}\label{dstau}
\om_j=\frac{\pd\log\tau}{\pd t_j}, \quad j\in E_+.
\end{equation}
It follows from Lemma~\ref{thm-UN} that $\log\tau$ is independent of
the choice of gauge equivalence slice of $\sL$.

\begin{dfn}
The function $\tau$ satisfying \eqref{dstau} is called tau function
of the Drinfeld-Sokolov hierarchy \eqref{Lt2}.
\end{dfn}

Let us consider how the tau function is related to the densities of
Hamiltonians in Theorem~\ref{thm-dsham}. The Hamiltonian densities
are
\begin{equation}\label{sHj2}
h_j=(-\Ld_j\mid H), \quad j\in E_+
\end{equation}
with $H$ given in Proposition~\ref{thm-dr}, and they are invariant
with respect to gauge transformations. Recall the definition of
$\om_j$ in \eqref{OmH}, one has
\begin{equation}\label{omhj}
\frac1{j}\,\frac{\pd\om_j}{\pd x}=\frac{(\Ld_j\mid -H)}{(\Ld_j\mid
\Ld_{-j}) }=\frac{h_j}{(\Ld_j\mid \Ld_{-j}) }.
\end{equation}
This together with \eqref{omtij} leads to
\begin{equation}
\frac{j}{(\Ld_j\mid \Ld_{-j})}\frac{\pd h_j}{\pd
t_i}=\frac{i}{(\Ld_i\mid \Ld_{-i})}\frac{\pd h_i}{\pd t_j}, \quad
i,j\in E_+.
\end{equation}
It means that a family of tau-symmetric Hamiltonian densities of the
Drinfeld-Sokolov hierarchy is found.

Equation \eqref{omhj} can be written as
\begin{equation}\label{DStau}
\frac{\pd^2 \log\tau}{\pd x\,\pd
t_j}=\frac{j}{(\Ld_j\mid\Ld_{-j})}h_j, \quad j\in E_+,
\end{equation}
which is just \eqref{DStau0}. Hence the tau function of the
Drinfeld-Sokolov hierarchy can be defined equivalently by
\begin{equation}\label{Htau}
\frac{\pd^2 \log\tau}{\pd t_i\,\pd
t_j}=\frac{j}{(\Ld_j\mid\Ld_{-j})}\pd_x^{-1}\left(-\Ld_j\mid
\frac{\pd H}{\pd t_i}\right), \quad i, j\in E_+.
\end{equation}
The right hand side is independent of the choice of nondegenerate
invariant symmetric bilinear form, in which $\pd H/\pd t_i$ is a
total derivative (see \eqref{HG}), and the integral constant is
assumed to be zero. Thus $\log\tau$ is determined up to the addition
of a linear function of the time variables.

It is natural to ask what is the relation between the above tau
function and those tau functions given in the literature. This
question will be dealt with below in Section~\ref{sec-exa} and the
appendix, for the reason that more notations are needed there.

\subsection{Zero-curvature representation for Drinfeld-Sokolov hierarchies}

We want to represent the Drinfeld-Sokolov hierarchies in a
zero-curvature form, which will be applied in the next section.

Given an operator $\sL$ as \eqref{sLgp}, the functions $U$ in
Proposition~\ref{thm-dr} and $\Om$ in \eqref{OmH} take value in
$\fg'^{<0}$, then the following element of the Lie group of $\fg'$
is well defined
\begin{equation}\label{Ta0}
\Ta=e^U e^{\Om}.
\end{equation}
This can be considered as a formal series that converges with
respect to a topology induced by the principal gradation on $\fg'$.
Note that generally the function $\Ta$ may not be a differential
polynomial in $q$.

Recalling $\Ld_1=\nu\Ld$ with constant $\nu$, we have
\begin{equation}\label{}
\sL=e^{\ad_U}e^{\ad_\Om}\left(D+\Ld+\frac{\om_1}{\nu}c\right)
=\Ta\left(D+\Ld+\frac{\om_1}{\nu}c\right)\Ta^{-1}.
\end{equation}
Note that $\Ta$ is determined by $\sL$ up to multiplication to the
right by $\exp\left(\sum_{j\in E_+}c_j\Ld_{-j}\right)$ with
constants $c_j$. The gauge transformation \eqref{gauge} of $\sL$
induces a transformation of the dressing operator as $\Ta\mapsto
e^N\Ta$. Hence there is a gauge slice of $\sL$ such that the
dressing operator takes the form (see \cite{HM})
\begin{equation}\label{Ta}
\Ta=e^V, \quad V\in C^\infty(\R, \fg'_{<0}).
\end{equation}
In the sequel we will fix such a gauge slice.

Note $\pd/\pd t_1=\nu\,\pd/\pd x$. We let
\begin{equation}\label{sL1}
\sL_1=\Ta\left(\frac{\pd}{\pd t_1}+\Ld_1\right)\Ta^{-1},
\end{equation}
namely, $\sL_1=\nu\sL-\om_1\,c$.

\begin{lem}\label{thm-LL1}
The evolutionary equations in \eqref{Lt2} are equivalent to
\begin{equation}\label{Lt}
\frac{\pd\sL_1}{\pd t_j}=[-(\Ta\Ld_j\Ta^{-1})_{\ge0},\sL_1],\quad
j\in E_+.
\end{equation}
\end{lem}
\begin{prf}
Since $\Ta\Ld_j\Ta^{-1}=\varphi(\Ld_j)-\om_j\,c$, then the
off-center part of \eqref{Lt} coincides with \eqref{Lt2}, while the
center part is just \eqref{om1tj} that can be derived from
\eqref{Lt2}. Thus the lemma is proved.
\end{prf}

The result of the following lemma have existed in \cite{HMSG, Mi}.
\begin{lem}\label{thm-Tat}
The dressing operator $\Ta$ in \eqref{Ta} satisfies
\begin{equation}\label{Tat}
\frac{\pd\Ta}{\pd t_j}=(\Ta\Ld_j\Ta^{-1})_{<0}\,\Ta, \quad j\in E_+.
\end{equation}
\end{lem}
\begin{prf}
Equations \eqref{Lt} can be written as
\begin{equation}
\frac{\pd\sL_1}{\pd t_j}=[(\Ta\Ld_j\Ta^{-1})_{<0},\sL_1],\quad j\in
E_+.
\end{equation}
Substitute into it with \eqref{sL1}, then one has
\[
\left[\frac{\pd\Ta}{\pd t_j}\Ta^{-1}-(\Ta\Ld_j\Ta^{-1})_{<0},
\sL_1\right]=0,\quad j\in E_+.
\]
For $j\in E_+$, denote
\begin{equation}\label{}
\Dt(j)=\Ta^{-1}\frac{\pd\Ta}{\pd t_j}
-\Ta^{-1}(\Ta\Ld_j\Ta^{-1})_{<0}\Ta,
\end{equation}
then $\Dt(j)\in C^\infty(\R,\fg'_{<0})$ and
\begin{equation}\label{Dtj}
\left[\Dt(j),\frac{\pd}{\pd t_1}+\Ld_1\right]=0.
\end{equation}
According to the decomposition \eqref{dec2}, we derive
\begin{equation*}
\Dt(j)=\sum_{i\in E_+}c_i\,\Ld_{-i}
\end{equation*}
with some constants $c_i$. But equation \eqref{Dtj} is independent
of $q$, hence by letting $q=0$ (which implies $U=0$ and $\Om=0$) one
obtains $\Dt(j)=0$. Thus we arrive at \eqref{Tat} and conclude the
lemma.
\end{prf}

Conversely, starting from \eqref{Tat} it is easy to derive equations
\eqref{Lt}, which yields the Drinfeld-Sokolov hierarchy \eqref{Lt2}.

With the operator $\Ta$ in \eqref{Ta}, let us introduce
\begin{equation}\label{}
\sL_j=\Ta\left(\frac{\pd}{\pd t_j}+\Ld_j\right)\Ta^{-1}, \quad j\in
E_+.
\end{equation}
By using Lemma~\ref{thm-Tat}, one can write these operators as
\begin{equation}\label{sLj}
\sL_j=\frac{\pd}{\pd t_j}+\Ld_j+q(j),\quad j\in E_+,
\end{equation}
where $q(j)=(\Ta\Ld_j\Ta^{-1})_{\ge0}-\Ld_j$. Thanks to \eqref{Ta0}
and \eqref{ULdc}, one sees
\begin{equation}\label{}
q(j)_c=-\om_j, \quad j\in E_+.
\end{equation}

Clearly, the operators \eqref{sLtj} satisfy
\begin{equation}\label{sLij}
[\sL_i,\sL_j]=0, \quad i,j\in E_+.
\end{equation}
This gives the zero-curvature representation for the
Drinfeld-Sokolov hierarchy \eqref{Lt}. In fact, it also confirms the
commutativity between the flows in \eqref{Lt}.



\section{Virasoro symmetries} \label{sec-vir}

It was shown \cite{HMSG} that the (generalized) Drinfeld-Sokolov
hierarchy associated to an untwisted affine Kac-Moody algebra
possesses certain additional symmetries that obey a Virasoro
commutation relation. Now we want to revise the construction in
\cite{HMSG} and derive such Virasoro symmetries for Drinfeld-Sokolov
hierarchy associated to an arbitrary affine Kac-Moody algebra. Our
aim is to represent these Virasoro symmetries via the tau function
defined in the previous section.

\subsection{Kac-Moody-Virasoro algebras and their representations}
Based on \cite{Kac, Wa}, we review the extension of the affine
algebra $\fg(A)$ to a Kac-Moody-Virasoro algebra, as well as some
properties of their representations. Suppose the Cartan matrix
$A=(a_{i j})_{0\le i, j\le n}$ is of affine type $X_N^{(r)}$; the
lowest positive integers $k_i$ satisfying $\sum_{j=0}^n a_{i
j}k_j=0$ are called the Kac labels of $\fg(A)$. The set of
gradations on $\fg(A)$ is
\begin{equation}\label{}
\Gm=\{(s_0, s_1, \dots, s_n)\in\Z^{n+1}\mid s_i\ge0,
s_0+s_1+\cdots+s_n>0\}.
\end{equation}
For every $\rs=(s_0, s_1, \dots, s_n)\in\Gm$, denote
\begin{equation}\label{}
N_{\rs}=\sum_{i=0}^n k_i s_i.
\end{equation}
In particular, recalling the homogeneous and the principal
gradations \eqref{hmgr}--\eqref{prgr}, we have $N_{\rs^0}=k_0$, and
$N_{\rs^1}=h$ being the Coxeter number of $\fg(A)$.

Let $\mathcal{G}$ be the simple Lie algebra of type $X_N$, on which
there is a diagram automorphism of order $r$. Given an integer
vector $\rs=(s_0, s_1, \dots, s_n)\in\Gm$, it induces a $\Z/r
N_\rs\Z$-gradation $\mathcal{G}=\bigoplus_{k=0}^{r
N_\rs-1}\mathcal{G}_k$ (see \S\,8.6 of \cite{Kac} for details). The
Kac-Moody algebra $\fg(A)$ graded by $\rs$ can be realized as
\begin{equation}\label{gAs}
\fg(A;\rs)=\bigoplus_{k\in\Z}\left(\ld^k\otimes\mathcal{G}_{k\!\!\mod\,r
N_{\rs}}\right)\oplus \C\,c\oplus \C d_0^{(\rs)},
\end{equation}
in which $c$ is the canonical central element, and the Lie bracket
between $X(k), Y(k)\in \ld^k\otimes\mathcal{G}_{k\!\!\mod\,r
N_{\rs}}$ and $d_0^{(\rs)}$ is defined by
\begin{align}\label{XYbr}
&[X(k), Y(l)]=[X, Y](k+l)+\dt_{k,-l}\frac{k}{r N_\rs}(X\mid Y)_0\, c, \\
&[d_0^{(\rs)}, X(k)]=k\,X(k).  \label{d0X}
\end{align}
Here $(\,\cdot\mid\cdot\,)_0$ is the standard invariant symmetric
bilinear form on $\mathcal{G}$. An element $X(k)$ will be written
more precisely as $X(k;\rs)$ whenever it is necessary to distinguish
the gradation $\rs$ from others.

On the derived algebra  $\fg'(A;\rs)$ of $\fg(A;\rs)$ one introduces
a family of derivations $d_l^{(\rs)}~( l\in\Z)$ such that
\begin{align}\label{}
& [d_l^{(\rs)}, X(k)]=k X(k+r N_\rs l), \quad [d_l^{(\rs)}, c]=0,
\\
&  [d_k^{(\rs)}, d_l^{(\rs)}]=r N_\rs (l-k)  d_{k+l}^{(\rs)}.
\label{dskdsl}
\end{align}
These derivations generate an infinite-dimensional Lie algebra, say,
$\mathfrak{d}^{(\rs)}$, of Virasoro type. Thus a Kac-Moody-Virasoro
algebra $\mathfrak{d}^{(\rs)}\ltimes\fg'(A;\rs)$ is constructed.

Following the notations in \S\,8.3 of \cite{Kac}, the simple Lie
algebra $\mathcal{G}$ contains certain elements written as $E_i$,
$F_i$ and $H_i$ with $i=0, 1, \dots, n$. These elements give a set
of Weyl generators of $\fg'(A;\rs)$ as follows: for $i=0, 1, \dots,
n$,
\begin{equation}\label{weyls}
e_i^{(\rs)}=E_i(s_i), \quad  f_i^{(\rs)}=F_i(-s_i), \quad
\al_i^{\vee(\rs)}=H_i(0)+\frac{k_i s_i}{k^\vee_i N_\rs}c,
\end{equation}
with $k_i$ and $k^\vee_i$ being the Kac labels and the dual Kac
labels respectively. Hence $d_k^{(\rs)}$ can be considered as
derivations on $\fg'(A)$.

For two arbitrary gradations $\rs, \rs'\in\Gm$, there is a natural
isomorphism between $\fg'(A;\rs)$ and $\fg'(A;\rs')$ induced by
\[
e_i^{(\rs)}\mapsto e_i^{(\rs')}, \quad  f_i^{(\rs)}\mapsto
f_i^{(\rs')}, \qquad i=0, 1, \dots, n.
\]
Up to such an isomorphism, one has the following lemma.
\begin{lem}[\cite{Wa}]\label{thm-dkl}
Given two gradations $\rs, \rs'\in\Gm$, the corresponding
derivations on $\fg'(A)$ satisfy
\begin{align}\label{}
&d_k^{(\rs+\rs')}=d_k^{(\rs)}+d_k^{(\rs')}, \\
&[d_k^{(\rs)}, d_l^{(\rs')}]=r N_{\rs} l d_{k+l}^{(\rs')}-r N_{\rs'}
k d_{k+l}^{(\rs)}
\end{align}
for all integers $k$ and $l$.
\end{lem}

In the Cartan subalgebra of $\fg'(A;\rs)$, one introduces the
following elements
\begin{align}\label{hs}
&h_\rs=(\al_1^{\vee(\rs)}, \dots,
\al_n^{\vee(\rs)})\left(\mathring{A}^T\right)^{-1}(s_1, \dots,
s_n)^T, \\
&H_\rs=(H_1(0), \dots, H_n(0))\left(\mathring{A}^T\right)^{-1}(s_1,
\dots, s_n)^T, \label{Hs}
\end{align}
where $\mathring{A}=(a_{i j})_{1\le i, j\le n}$, and the superscript
``$T$'' means the transpose of matrices.

\begin{lem}[Lemma~2.4 in \cite{Wa}] \label{thm-dks}
Let $\rs^0=(1,0,\dots,0)\in\Gm$ be the homogeneous gradation. For
any $\rs\in\Gm$ it holds that
\begin{equation}\label{}
d_k^{(\rs)}=\left\{ \begin{array}{ll}
              \dfrac{N_\rs}{k_0}d_0^{(\rs^0)}+h_\rs, & k=0; \\
              \\
              \dfrac{N_\rs}{k_0}d_k^{(\rs^0)}+H_\rs(r k_0 k; \rs^0), & k\ne0.
            \end{array}
            \right.
\end{equation}
\end{lem}

\vskip 1ex

Consider highest weight representations of affine Kac-Moody
algebras. Denote $\bx=(x_j)_{j\in E_+}$, then on the Fock space
$\C[[\bx]]$ one defines an action of the principal Heisenberg
subalgebra $\mathfrak{s}$ of $\fg'(A)$ by
\begin{equation}\label{srep}
c\mapsto1; \quad \Ld_j\mapsto -\frac{\pd}{\pd x_j}, \quad
\Ld_{-j}\mapsto - j\, x_j, \quad j\in E_+.
\end{equation}
This action generates a highest weight representation of
$\mathfrak{s}$.
\begin{thm}[Theorem~14.6 in \cite{Kac}] \label{thm-kacb}
Suppose the Dynkin diagram of the affine Lie algebra $\fg'(A)$ is
simply-laced or twisted, then the action of the Heisenberg
subalgebra $\mathfrak{s}$ on the Fock space $\C[[\bx]]$ given by
\eqref{srep} can be lifted to a basic representation $L(\Ld_0)$ of
$\fg'(A)$ on $\C[[\bx]]$, and, the highest weight vector is $1$.
\end{thm}

Recall the principal gradation $\rs^1=(1,1,\dots,1)\in\Gm$, and
realize $\fg(A)$ as $\fg(A;\rs^1)$. The following result is implied
by Theorems~3.2 and 5.1 in \cite{Wa}.
\begin{thm}\label{thm-rep}
Under the same assumption as in Theorem~\ref{thm-kacb}, the action
of $\fg'(A)$ on $\C[[\bf{x}]]$ given there can be uniquely extended
to a $\mathfrak{d}_k^{(\rs^1)}\ltimes\fg'(A;\rs^1)$-action by
setting
\begin{equation}\label{}
d_k^{(\rs^1)}\mapsto -L_k(\pd/\pd\bx; \bx), \quad k\in \Z,
\end{equation}
where
\begin{equation}\label{virLk}
L_k(\pd/\pd\bx; \bx)=\frac{1}{2}\sum_{j\in E_+}\left(p_{r h
k-j}p_j+p_{-j}p_{r h k+j}\right)
\end{equation}
with $h=N_{\rs^1}$ being the Coxeter number, and
\begin{equation*}\label{}
p_j=\frac{\pd}{\pd x_j}, \quad p_{-j}=j\,x_j, \qquad j\in E_+.
\end{equation*}
\end{thm}

\subsection{Virasoro symmetries represented via tau function}

We proceed to construct Virasoro symmetries for the Drinfeld-Sokolov
hierarchy, and consider their action on the tau function defined in
the previous section.

Let $\fg$ be an arbitrary affine Kac-Moody algebra. For the
homogeneous and the principal gradations $\rs^0$ and $\rs^1$,
recalling  $N_{\rs^0}=k_0$ and $N_{\rs^1}=h$, we normalize the
derivations on $\fg'$ as:
\begin{align}\label{}
&d_k=-\frac1{r N_{\rs^0}}d_k^{(\rs^0)}=-\frac1{r k_0}d_k^{(\rs^0)},
\\
&d_k'=-\frac1{r N_{\rs^1}}d_k^{(\rs^1)}=-\frac1{r h}d_k^{(\rs^1)}.
\end{align}
Thanks to \eqref{dskdsl}, these normalized derivations satisfy
\begin{equation}\label{dkdl}
[d_k, d_l]=(k-l)d_{k+l}, \quad [d_k', d_l']=(k-l)d_{k+l}', \qquad
k,l\in\Z.
\end{equation}
From Lemma~\ref{thm-dks} it follows that
\begin{equation}\label{ddm}
d_k-d_k'=\frac{1}{r
h}\left(d_k^{(\rs^1)}-\frac{h}{k_0}d_k^{(\rs^0)}\right)
=\left\{\begin{array}{cl} \dfrac{1}{r h}h_{\rs^1}, & k=0; \\
\\
 \dfrac{1}{r h}H_{\rs^1}(r k_0 k;\rs^0), & k\ne0.\end{array} \right.
\end{equation}
Clearly $d_k-d_k'\in\fg'_{r k_0 k}$ in the notations for the
homogeneous gradation \eqref{hmgr}.

We fix the generators $\Ld_j\in\fg'^j$ with $j\in E$ of the
principle Heisenberg algebra $\mathfrak{s}$ such that
\begin{equation}\label{dLd}
[\Ld_i,\Ld_j]=\dt_{i,-j}\,i\cdot c, \quad  [d_k', \Ld_j]=-\frac{j}{r
h}\Ld_{j+r h k} ~\hbox{ for all }~ k\in\Z.
\end{equation}
Given any integer $k$, introduce
\begin{align}\label{Bk0}
\tilde{B}_k=d_k'-\sum_{i\in E_+}\frac{i t_i}{r h}\Ld_{i+r h
k}+\frac{1}{2 r h}\sum_{\scriptsize\hbox{$\begin{array}{c} i,j\in
E_+ \\ i+j=-r h k \end{array}$}} i j t_i t_j\cdot c,
\end{align}
where the third term on the right hand side exists only if $k<0$. It
is straightforward to check
\begin{equation}\label{Bkt}
\left[\tilde{B}_k, \frac{\pd}{\pd t_j}+\Ld_j\right]=0, \qquad j\in
E_+, \quad k\in\Z.
\end{equation}
Recalling the dressing operator $\Ta$ in \eqref{Ta}, we let
\begin{equation}\label{Bk}
B_k=\Ta\tilde{B}_k\Ta^{-1}-d_k, \quad k\in\Z.
\end{equation}
Note $B_k\in\fg'$ for all $k$, and that \eqref{Bkt} leads to
\begin{equation}\label{}
[B_k+d_k,\sL_j]=0,  \qquad j\in E_+, \quad  k\in\Z.
\end{equation}

For $k\ge-1$, we define a class of evolutionary equations as follows
\begin{equation}\label{Lbt}
\frac{\pd\sL_1}{\pd \beta_k}=\left[-(B_k)_{<0}, \sL_1\right].
\end{equation}
The right hand side can also be rewritten as
$\left[(B_k)_{\ge0}+d_k, \sL_1\right]$, hence the equations
\eqref{Lbt} are well defined by comparing the degrees with respect
to the homogeneous and the principal gradations, as well as by an
analysis like in the proof of Lemma~\ref{thm-LL1} for the central
part. The flows \eqref{Lbt} are assumed to commute with $\pd/\pd
t_1=\nu\,\pd/\pd x$.

With the same method as to prove Lemma~\ref{thm-Tat}, we have the
following useful formulae
\begin{equation}\label{Tabt}
\frac{\pd \Ta}{\pd \beta_k}=-(B_k)_{<0}\Ta, \quad k\ge-1.
\end{equation}
This formulae lead immediately to
\begin{equation}\label{Ldjbt}
\frac{\pd\sL_j}{\pd \beta_k}=\left[-(B_k)_{<0},
\sL_j\right]=\left[(B_k)_{\ge0}+d_k, \sL_j\right], \qquad k\ge-1,
\quad j\in E_+.
\end{equation}

\begin{prp}\label{thm-btt}
The following assertions are true (cf. Propositions~3.3 and 3.4 in
\cite{HMSG}):
\begin{itemize}
\item[(i)]
The flows \eqref{Lbt} commute with those in \eqref{Lt}, or
equivalently,
\begin{equation}\label{btt}
\left[\frac{\pd}{\pd\beta_k}, \frac{\pd}{\pd t_j}\right]\Ta=0, \quad
k\ge-1, ~~ j\in E_+.
\end{equation}
\item[(ii)] For $k, l\ge-1$,
\begin{equation}\label{btbt}
\left[\frac{\pd}{\pd\beta_k},
\frac{\pd}{\pd\beta_l}\right]\Ta=(l-k)\frac{\pd
\Ta}{\pd\beta_{k+l}}.
\end{equation}
\end{itemize}
\end{prp}
\begin{prf}
The first assertion follows from a straightforward calculation by
using \eqref{Tabt} and \eqref{Tat}. Let us check the second
assertion, which is more nontrivial.

Using \eqref{dkdl} and \eqref{dLd}, one can verify
\begin{align*}
&[\tilde{B}_k, \tilde{B}_l]=(k-l)\tilde{B}_{k+l}.
\end{align*}
Since
\begin{align*}
\frac{\pd}{\pd\beta_k}\frac{\pd}{\pd\beta_l}\Ta
=&-\frac{\pd}{\pd\beta_k}(B_l)_{<0}\Ta \\
=&(B_l)_{<0}(B_k)_{<0}\Ta
+[(B_k)_{<0},\Ta\tilde{B}_l\Ta^{-1}]_{<0}\Ta\\
=&(B_l)_{<0}(B_k)_{<0}\Ta +[(B_k)_{<0},B_l+d_l]_{<0}\Ta,
\end{align*}
then
\begin{align}
&\left[\frac{\pd}{\pd\beta_k},\frac{\pd}{\pd\beta_l}\right]\Ta\cdot\Ta^{-1}
\nn\\
=&\left[(B_l)_{<0}, (B_k)_{<0}\right]+ \left[(B_k)_{<0},
B_l+d_l\right]_{<0}- \left[(B_l)_{<0}, B_k+d_k\right]_{<0} \nn\\
=& [B_k, B_l]_{<0}+[(B_k)_{<0},d_l]_{<0}+[d_k, (B_l)_{<0}]_{<0}
\nn\\
=& \left( [B_k+d_k, B_l+d_l]-[d_k,d_l]\right)_{<0}
\nn\\
=&\left((k-l)(B_{k+l}+d_{k+l})-(k-l)d_{k+l}\right)_{<0}
\nn \\
=&(k-l)(B_{k+l})_{<0}
\nn \\
=&(l-k)\frac{\pd\Ta}{\pd\beta_{k+l}}\Ta^{-1}.
\end{align}
Note that in the third equality we have used $[d_{k},
B_{\ge0}]_{<0}=0$ for any element $B\in\fg'$ and integer $k\ge-1$.
The proposition is proved.
\end{prf}

The commutativity \eqref{btt} of flows means that equations
\eqref{Bkt} define symmetries for the Drinfeld-Sokolov hierarchy
\eqref{Lt2}. Such symmetries are called the \emph{Virasoro
symmetries} due to the commutation relation \eqref{btbt}.

Now we arrive at the main result of the present section.
\begin{thm}\label{thm-taubt}
For $k\ge-1$, the tau function in \eqref{dstau} of the
Drinfeld-Sokolov hierarchy satisfies
\begin{align}
\frac{\pd\log\tau}{\pd\beta_k}=&\frac{1_{k\ge1}}{r h}O_k'+
\frac{1}{2 r h}\sum_{i+j=r h
k}\frac{\pd\log\tau}{\pd t_i}\frac{\pd\log\tau}{\pd t_j} \nn\\
&+\frac{1}{r h}\sum_{i>-r h k}i t_i\frac{\pd\log\tau}{\pd t_{i+r h
k}}+\frac{1}{2 r h}\sum_{i+j=-r h k}i j t_i t_j + \dt_{k,0}{C_A}.
\label{taubt}
\end{align}
Here $1_{k\ge1}$ equals $1$ whenever $k\ge1$ and vanishes otherwise,
\begin{equation}\label{Jk}
O_k'=r h \left(e^{\ad_U}d_k-d_k\right)_c- \left(e^{\ad_U}H_{\rs^1}(r
k_0 k;\rs^0)\right)_c
\end{equation}
with $U$ given in Proposition~\ref{thm-dr} and $H_{\rs^1}(r k_0
k;\rs^0)$ introduced from \eqref{Hs}, $C_A$ is a constant depending
on the Cartan matrix $A$, and all indices $i, j\in E_+$. Note that
$O_k'$ are independent of the gauge transformations \eqref{gauge}.
\end{thm}
\begin{prf}
Recall \eqref{sLj}. The equations in \eqref{Ldjbt} yield
\begin{equation}\label{qBc}
\frac{\pd q(j)_c}{\pd\beta_k}=-\frac{\pd (B_k)_c}{\pd t_j}, \quad
k\ge-1, j\in E_+.
\end{equation}
The left hand side is
\begin{equation}\label{taubt1}
-\frac{\pd\om_j}{\pd\beta_k}=-\frac{\pd^2\log\tau}{\pd\beta_k\pd
t_j};
\end{equation}
let us compute $(B_k)_c$ on the right hand side of \eqref{qBc}.

Recalling  \eqref{OmH} and \eqref{dLd}, it is straightforward to
calculate
\begin{align}
e^{\ad_\Om}\tilde{B}_k =&e^{\ad_\Om}\left(d_k'-\sum_{i\in
E_+}\frac{i t_i}{r h}\Ld_{i+r h k}+\frac{1}{2 r h}\sum_{ i+j=-r h k}
i j t_i t_j\cdot c\right)
\nn\\
=&d_k'-\frac{1}{r h}\sum_{i\in E_+}\om_i \Ld_{-i+r h k}+\frac{1}{2 r
h}\sum_{i+j=r h k}\om_i\om_j\cdot c \nn\\
&-\sum_{i\in E_+}\frac{i t_i}{r h}\Ld_{i+r h k}-\sum_{i>-r h
k}\frac{i t_i}{r h}\frac{\om_{i+r h k}}{i+r h k}(-i-r h k)\cdot
c\nn\\
&+\frac{1}{2 r h}\sum_{ i+j=-r h k} i j t_i t_j\cdot c \nn\\
=&d_k'-\frac{1}{r h}\sum_{i\in E_+}(\om_i\Ld_{-i+r h k} + i
t_i\Ld_{i+r h k}) +\frac{1}{2 r h}\sum_{i+j=r h k}\om_i\om_j\cdot c
\nn \\
&+\frac{1}{r h}\sum_{i>-r h k}i\, t_i\,\om_{i+r h k}\cdot
c+\frac{1}{2 r h}\sum_{ i+j=-r h k} i j t_i t_j\cdot c.
\end{align}
By virtue of the property of $U$ given in Proposition~\ref{thm-dr},
we arrive at
\begin{align}
(B_k)_c=& \left(e^{\ad_U}e^{\ad_\Om}\tilde{B}_k-d_k\right)_c \nn\\
=&\left(e^{\ad_U}d_k'-d_k\right)_c+ \frac{1}{2 r h}\sum_{i+j=r h
k}\om_i\om_j \nn\\
&+\frac{1}{r h}\sum_{i>-r h k}i\, t_i\,\om_{i+r h k}+\frac{1}{2 r
h}\sum_{ i+j=-r h k} i j t_i t_j.\label{Bkc}
\end{align}
Here we have used the fact that $(B_k)_c$ is independent of  the
gauge transformations \eqref{gauge}, for the same reason as in the
proof of Lemma~\ref{thm-UN}.

Denote $O_k'=r h\left(e^{\ad_U}d_k'-d_k\right)_c$. Clearly,
\begin{equation}
O_{-1}'=0, \quad O_0'=r h(d_0'-d_0)_c=-\left(h_{\rs^1}\right)_c=0.
\end{equation}
When $k\ge1$, by using \eqref{ddm} we have
\begin{align}
O_k'=&r h\left(e^{\ad_U}\left(d_k-\frac{1}{r h}H_{\rs^1}(r k_0
k;\rs^0)\right)-d_k \right)_c \nn\\
=&r h \left(e^{\ad_U}d_k-d_k\right)_c- \left(e^{\ad_U}H_{\rs^1}(r
k_0 k;\rs^0)\right)_c. \label{J1}
\end{align}
We substitute \eqref{taubt1}, \eqref{Bkc}--\eqref{J1} into
\eqref{qBc}, then obtain \eqref{taubt} by integration with respect
to $t_j$. The constant $C_A$ is chosen such that $\pd/\pd\beta_k$
acting on $\tau$ obey the Virasoro commutation relation. The theorem
is proved.
\end{prf}

In general, we only know that $O_k'$ are differential polynomials in
second-order derivatives of $\log\tau$ with respect to the time
variables $t_j$. The reason is that in Proposition~\ref{thm-dr} the
function $U$ is a differential polynomial in $q$, which can be
chosen canonically as a differential polynomial in the first $n$
Hamiltonian densities $h_j=(-\Ld_j\mid H)$ with exponents $j$ such
that $0<j<r h$. How to compute $O_k'$ will be illustrated by
examples in next section.

We continue to write \eqref{taubt} into a more concise form. Observe
that the Sugawara construction of $L_k$ in \eqref{virLk} can be
extended to an arbitrary affine Kac-moody algebra $\fg$. Let
\begin{equation}\label{Vk}
V_k=\frac{1}{r h}L_k(\pd/\pd\bt; \bt)+\dt_{k,0}\cdot C_A,
\end{equation}
and they satisfy
\begin{equation*}\label{}
[V_k, V_l]=(k-l)V_{k+l}, \quad k, l\ge-1.
\end{equation*}
Theorem~\ref{thm-taubt} leads us to
\begin{cor}\label{thm-Vtau}
The Virasoro symmetries \eqref{Lbt} for the Drinfeld-Sokolov
hierarchies can be represented via tau function as
\begin{equation}\label{tauvir}
\frac{\pd\tau}{\pd\beta_k}=V_k\tau+\tau\,O_k, \quad k\ge-1,
\end{equation}
where
\begin{equation}\label{Ok}
O_k=\left\{ \begin{aligned}
              &0, && k=-1, 0; \\
              & & & \\
              &\frac{1}{r h}\left(O_k'-\frac{1}{2}\sum_{i+j=r h k}
\frac{\pd^2\log\tau}{\pd t_i\,\pd t_j}\right), & & k\ge1.
            \end{aligned}
\right.
\end{equation}
\end{cor}

In particular, the symmetries generated by $\pd/\pd\beta_{-1}$ and
$\pd/\pd\beta_{0}$ are linearized when acting on the tau function.
In the next subsection we will show that, when the Drinfeld-Sokolo
hierarchies are associated to ADE-type affine Kac-Moody algebras,
the symmetries generated by $\pd/\pd\beta_{k}$ with all $k\ge-1$ are
linearized. However, such kind of linearization of Virasoro
symmetries is not valid for all Drinfeld-Sokolov hierarchies due to
the functions $O_k$ may not vanish.

\begin{dfn}
The functions $O_k$ in \eqref{tauvir} are called obstacles in
linearizing Virasoro symmetries.
\end{dfn}

\subsection{Linearization of Virasoro symmetries for ADE-type hierarchies}

In this subsection we only consider the Drinfeld-Sokolov hierarchies
associated to simply-laced or twisted affine Kac-Moody algebras, for
which the Virasoro symmetries acting on the tau function will be
seen linearized.

\begin{lem}\label{thm-Tatau}
Let $\fg'$ be a simply-laced or twisted affine Lie algebra. The
basic representation given in Theorem~\ref{thm-kacb} induces an
action of the Lie group on the Fock space $\C[[\bx]]$, then the
operator $\Ta$ introduced in \eqref{Ta} satisfies
\begin{equation}\label{Tatau}
\Ta^{-1}\cdot1=\frac{\tau(\bt+\bx)}{\tau(\bt)},
\end{equation}
where $\tau$ is the tau function of the corresponding
Drinfeld-Sokolov hierarchy.
\end{lem}
\begin{prf}
In the basic representation $\C[[\bx]]$ of $\fg'$, every element
$X\in\fg'_{\ge0}$ satisfies $X\cdot1=X_c$. For each positive
exponent $j\in E_+$,
\begin{align}
\left(\frac{\pd}{\pd
t_j}+\Ld_j\right)\Ta^{-1}\cdot1&=\Ta^{-1}\sL_j\cdot1 \nn\\
&=\Ta^{-1}\left(\frac{\pd}{\pd
t_j}+q(j)+\Ld_j\right)\cdot1 \nn\\
&=\Ta^{-1} q(j)_c\,c\cdot1 \nn\\
&= -\om_j\,\Ta^{-1}\cdot1.
\end{align}
Denote $\Ta^{-1}\cdot1=f(\bt, \bx)$, then the above equation is just
\begin{align}
\left(\frac{\pd}{\pd t_j}-\frac{\pd}{\pd x_j}\right)f(\bt,
\bx)=-\frac{\pd \log\tau(\bt)}{\pd t_j}f(\bt, \bx).
\end{align}
This equation together with the ``boundary'' condition
$f(\bt,0)=\big(\Ta^{-1}\cdot1\big)|_{\bx=0}=1$ implies $f(\bt,
\bx)={\tau(\bt+\bx)}/{\tau(\bt)}$. The lemma is proved.
\end{prf}

\begin{prp} \label{thm-virADE}
Suppose the affine Lie algebra $\fg'$ is simply-laced or twisted,
then the Virasoro symmetries \eqref{Lbt} for the Drinfeld-Sokolov
hierarchy act linearly on the tau function. More precisely,
\begin{equation}\label{Vktau}
\frac{\pd\tau}{\pd\beta_k}=V_k\tau, \quad k\ge-1,
\end{equation}
where the operators $V_k$ are given in \eqref{Vk}.
\end{prp}
\begin{prf}
We only need to show the cases $k\ge1$. The proof is almost the same
as that of Proposition~4.2 in \cite{HMSG} for simply-laced affine
Lie algebras.

Consider the basic representation $\C[[\bx]]$ of $\fg'$ as in
Theorem~\ref{thm-kacb}. It follows from the formula \eqref{Tatau}
that
\begin{equation}\label{Tainbt}
\frac{\pd}{\pd\beta_k}\Ta^{-1}\cdot1
=\frac{\pd}{\pd\beta_k}\frac{\tau(\bt+\bx)}{\tau(\bt)}
=\frac{1}{\tau(\bt)}\frac{\pd\tau(\bt+\bx)}{\pd\beta_k}
-\frac{\tau(\bt+\bx)}{\tau(\bt)^2}\frac{\pd\tau(\bt)}{\pd\beta_k}.
\end{equation}
On the other hand, let us extend the basic representation
$\C[[\bx]]$ to a module of the Kac-Moody-Virasoro algebra as in
Theorem~\ref{thm-rep}. Recall $p_i$ given in \eqref{srep}, and set
\[
\Ld_j\mapsto -p_j, ~~~ j\in E;  \quad  d_k'\mapsto\frac{1}{r
h}L_k(\pd/\pd\bx;\bx), ~~~ k\in\Z.
\]
When $k\ge1$, by using \eqref{Tabt} and \eqref{qBc} we have
\begin{align}\label{}
&\frac{\pd}{\pd\beta_k}\Ta^{-1}\cdot1 \nn\\
=&
-\Ta^{-1}\frac{\pd\Ta}{\pd\beta_k}\Ta^{-1}\cdot1=\Ta^{-1}(B_k)_{<0}\cdot1
\nn\\
=&\Ta^{-1}B_k\cdot1-\Ta^{-1}(B_k)_{\ge0}\cdot1 \nn\\
=&\tilde{B}_k\Ta^{-1}\cdot1 -\Ta^{-1}d_k\cdot1-(B_k)_c\Ta^{-1}\cdot1 \nn\\
=&\left(\frac{1}{r h}L_k(\pd/\pd\bx;\bx)+\frac{1}{r h}\sum_{j\in
E_+}j t_j p_{j+r h k} \right)\frac{\tau(\bt+\bx)}{\tau(\bt)} -0
\nn\\
&-\frac{\pd\log\tau(\bt)}{\pd\beta_k}
\frac{\tau(\bt+\bx)}{\tau(\bt)} \nn\\
=&\frac{1}{\tau(\bt)}\frac{1}{r
h}L_k(\pd/\pd\bx;\bt+\bx)\tau(\bt+\bx)-
\frac{\tau(\bt+\bx)}{\tau(\bt)^2}\frac{\pd\tau(\bt)}{\pd\beta_k}.
\label{Tainbt2}
\end{align}
Taking \eqref{Tainbt} and \eqref{Tainbt2} together, we obtain
\[
\frac{\pd\tau(\bt+\bx)}{\pd\beta_k}=\frac{1}{r
h}L_k(\pd/\pd\bx;\bt+\bx)\tau(\bt+\bx), \quad k\ge1,
\]
which is recast to \eqref{Vktau} by a translation of variables:
$\bt+\bx\mapsto\bt$. The proposition is proved.
\end{prf}

Comparing equations \eqref{taubt} and \eqref{Vktau}, we have
immediately
\begin{cor}\label{thm-okADE}
For any simply-laced or twisted affine Lie algebra, the obstacles
$O_k$ vanish for all $k\ge-1$. In other words, the functions $O_k'$
defined in \eqref{Jk} can be written as
\begin{equation}\label{JkADE}
O_k'=\frac{1}{2}\sum_{i+j=r h k} \frac{\pd^2\log\tau}{\pd t_i\,\pd
t_j}, \quad k\ge1.
\end{equation}
\end{cor}

\begin{prfof}{Theorem~\ref{thm-main}}
The theorem follows from Corollary~\ref{thm-Vtau} and
Proposition~\ref{thm-virADE}.
\end{prfof}

\vskip 2ex

At the end this section, we digress to consider another important
application of Lemma~\ref{thm-Tatau}. Recall that Kac and Wakimoto
\cite{KW} constructed a hierarchy of Hirota bilinear equations based
on the principal vertex operator realization of the basic
representation of $\fg'$. Every solution of these equations is a
point of the orbit space of the highest weight vector acted by the
Lie group. More exactly, let $L(\Ld_0)=\C[[\bx]]$ be the basic
representation of $\fg'$ given in Theorem~\ref{thm-kacb}, and $G$ be
the Lie group of $\fg'$, then $\tau\in L(\Ld_0)$ lies in the orbit
$G\cdot 1$ if and only if it satisfies a hierarchy of Hirota
bilinear equations of the form
\begin{equation}\label{}
R_\sg\,\tau\cdot\tau=0, \quad \sg\in S_A,
\end{equation}
where $R_\sg$ are certain constant-coefficient even polynomials in
$\{D_j\mid j\in E_+\}$ with Hirota differential operators $D_j$
given by
\[
D_j f\cdot g=\left.\frac{\pd}{\pd y}\right|_{y=0}f(x_j+y)g(x_j-y).
\]
See \cite{Kac, KW} for more details.

According to \cite{DS, JM83, KW}, it is known that the Kac-Wakimoto
hierarchies of bilinear equations for $\fg'$ of type $A_n^{(1)}$ is
equivalent with the corresponding Drinfeld-Sokolov hierarchies. Such
an equivalence was proved by us in \cite{LWZ} for the case of type
$D_n^{(1)}$, see also \cite{Wu}. In general, we have the following
\begin{thm}\label{thm-KW}
For any simply-laced or twisted affine Lie algebra $\fg'$, the tau
functions defined in \eqref{dstau} of the Drinfeld-Sokolov hierarchy
coincide with the solutions of the corresponding Kac-Wakimoto
hierarchy. The variables of these two hierarchies are related via a
basis of the principal subalgebra $\mathfrak{s}$ according to the
relations \eqref{Lt2} and \eqref{srep}.
\end{thm}
\begin{prf}
Given a tau function $\tau(\bt)$ of the Drinfeld-Sokolov hierarchy,
the formula \eqref{Tatau} implies that $\tau(\bt+\bx)/\tau(\bt)$ is
a solution of the Kac-Wakimoto hierarchy of Hirota equations with
variable $\bx$. By setting $\bt\to0$ one sees that $\tau(\bx)$ also
solves the Hirota equations due to their bilinearity and translation
invariance with respect to $\bx$.

Conversely, suppose $\tau(\bx)$ is a solution of the Kac-Wakimoto
hierarchy. The substitution of $\tau$ into the right hand side of
\eqref{Tatau} determines an element $\Ta=e^{V(\bt)}$ with $V(\bt)$
being a smooth function that takes value in $\fg'_{<0}$. For every
$j\in E_+$, introduce $\sL_j=\Ta\left(\pd/\pd
t_j+\Ld_j\right)\Ta^{-1}$. They act on the highest weight vector as
\begin{align}\label{}
\sL_j\cdot1=&\Ta\left(\frac{\pd}{\pd t_j}-\frac{\pd}{\pd
x_j}\right)\frac{\tau(\bt+\bx)}{\tau(\bt)}\cdot1 \nn\\
=&-\frac{\pd\log\tau(\bt)}{\pd
t_j}\frac{\tau(\bt+\bx)}{\tau(\bt)}\Ta\cdot1 \nn\\
=&-\frac{\pd\log\tau(\bt)}{\pd t_j},
\end{align}
which is independent of $\bx$. It implies that $\sL_j=\pd/\pd
t_j+\Ld_j+q(j)$, where $q(j)$ lies in $\fg'_{\ge0}$ and
\[
q(j)_c=-\frac{\pd\log\tau(\bt)}{\pd t_j}.
\]
The following zero-curvature equations
\begin{equation}\label{LiLj}
 [\sL_i, \sL_j]=0, \quad i,j\in E_+
\end{equation}
are well defined. These equations recovers the Drinfeld-Sokolov
hierarchy, of which the tau function defined by \eqref{dstau} is
just $\tau(\bt)$. The theorem is proved.
\end{prf}

\begin{rmk}
The results in the present subsection rely on the property of the
basic representation of the affine Lie algebra $\fg'$; they may not
be valid for Drinfeld-Sokolov hierarchies associated to other than
the zeroth vertex of the Dynkin diagram of $\fg'$. For example, the
Virasoro symmetries for the $A_2^{(2)}$-hierarchy associated to the
vertex labeled $1$ of the Dynkin diagram are not linearized; note
that the first non-trivial equation in this hierarchy is also known
as the Kaup-Kupershmidt equation \cite{Ka}.
\end{rmk}

Finally, we shall emphasize our inspiration from \cite{HM, HMSG}. In
fact, for each simply-laced affine Kac-Moody algebra $\fg'$,
Hollowood and Miramontes proved the following equality (see
equation~(5.1) in \cite{HM}):
\begin{equation}\label{hmtau}
\Ta^{-1}\cdot v_\rs=\frac{\tau_\rs(\bt+\bx)}{\tau_\rs^{(0)}(\bt)},
\end{equation}
with a method of ``big cell'' factorization of the Lie group of
$\fg'$. Here $\Ta$ is a dressing operator lying in the Lie subgroup
$U_-(\rs)$, $v_\rs$ is the highest weight vector in an integrable
highest weight representation $L(\rs)$, and $\tau_\rs$ is the tau
function of Hirota equations from Kac and Wakimoto's construction
\cite{KW}. The formula \eqref{hmtau} provides a map from solutions
of a Kac-Wakimoto hierarchy to those of the zero-curvature hierarchy
\eqref{LiLj} of Drinfeld-Sokolov type. 
This formula
was further employed in \cite{HMSG} to obtain the linearized
Virasoro symmetries for generalized Drinfeld-Sokolov hierarchies
associated to simply-laced affine Lie algebras, whose tau function
is considered to be $\tau_\rs$.

Inspired by Hollowood, Miramontes and S\'anchez Guill\'en \cite{HM,
HMSG}, we now obtain, in a more straightforward way, the formula
\eqref{Tatau} of the form \eqref{hmtau} (note the different
definitions of tau functions) whenever the affine algebra $\fg'$ is
simply-laced or twisted. For such cases, moreover, we clarify in
Theorem~\ref{thm-KW} the equivalence between the Drinfeld-Sokolov
and the Kac-Wakimoto hierarchies.

\subsection{Virasoro constraints to tau function}

Generally speaking, the tau function that gives partition function
in topological field theory is selected by the string equation.
Based on the Lax representation in pseudo-differential operators for
Drinfeld-Sokolov hierarchies of type $A_n^{(1)}$ or $D_n^{(1)}$, it
was shown that the string equation induces a series of Virasoro
constraints to the tau function \cite{AvM, Wu-vir}. Such kind of
constraints to the tau function $\tau_\rs$ in \eqref{hmtau} of any
Drinfeld-Sokolov hierarchy associated to simply-laced affine
Kac-Moody algebra $\fg$ were derived in \cite{HMSG}. It is not hard
to generalize the deduction in \cite{HMSG} to all Drinfeld-Sokolov
hierarchies and derive Virasoro constraints to the tau function
$\tau$ in \eqref{DStau}.

For the Drinfeld-Sokolov hierarchy associated to an arbitrary affine
Kac-Moody algebra $\fg$, we assume its tau function $\tau$ satisfies
the following string equation
\begin{equation}\label{}
\frac{\pd\tau}{\pd t_1}=V_{-1}\tau.
\end{equation}
This equation is just $\pd\tau/\pd t_1=\pd\tau/\pd \beta_{-1}$,
hence it follows from \eqref{Tat} and \eqref{Tabt} that
\begin{equation}\label{}
\left(\Ta\Ld_1\Ta^{-1}\right)_{<0}=-(B_{-1})_{<0}.
\end{equation}
Denote
\begin{equation}\label{}
P_k=\Ta\Ld_{1+r h (k+1)}\Ta^{-1}+B_k, \quad k\ge-1.
\end{equation}
First of all, one has $(P_{-1})_{<0}=0$.

Consider the realization of $\fg'$ graded by the homogeneous
gradation $\rs^0$, which is acted by a series of derivations
$d_k^{(\rs^0)}=\ld^{1+r k_0 k}\cdot{\od}/{\od\ld}$. Note that the
generators of the principal Heisenberg subalgebra $\mathfrak{s}$
satisfy
\[
\Ld_{j+r h(k+1)}=\ld^{r k_0}\Ld_{j+r h k},
\]
then recalling \eqref{Bk0} and \eqref{Bk} we have $P_{k+1}=\ld^{r
k_0} P_k$ modulo the central part. Hence we obtain
\begin{equation}\label{}
(P_k)_{<0}=(\ld^{(k+1)k_0}
P_{-1})_{<0}=\ld^{(k+1)k_0}(P_{-1})_{<-(k+1)k_0}=0, \quad k\ge-1.
\end{equation}
Therefore ${\pd\tau}/{\pd t_{1+r h(k+1)} }={\pd\tau}/{\pd
\beta_{k}}$, namely,
\begin{equation}\label{}
\frac{\pd\tau}{\pd t_{1+r h(k+1)} }=V_k\tau+\tau O_k, \quad k\ge-1.
\end{equation}
They are the Virasoro constraints to the tau function of the
Drinfeld-Sokolov hierarchy. We plan to study solutions to these
constraints elsewhere.

\begin{rmk}
In the recent paper \cite{Sa}, Safronov proposed a set of linear
Virasoro constraints to his tau function on Drinfeld-Sokolov
Grassmannians for the case of simply-laced semisimple Lie groups.
His Virasoro operators are also from the Sugawara construction
\eqref{virLk} for Heisenberg subalgebras, and the string solutions
are described geometrically by principal bundles possessing
connections compatible with the Higgs field near infinity.
\end{rmk}

\section{Examples} \label{sec-exa}

Let us present some examples to illustrate our construction of tau
function of Drinfeld-Sokolov hierarchies and the obstacles in
linearizing their Virasoro symmetries.

In the examples below we will use a matrix realization of $\fg$ of
affine type $X_N^{(r)}$ corresponding to the homogeneous gradation
$\rs^0$ as in \eqref{gAs}--\eqref{d0X} (see \cite{Kac} or the
appendix of \cite{DS}). In \eqref{XYbr} the standard invariant
symmetric bilinear form $(\,\cdot\mid\cdot)_0$ on $\mathcal{G}$ is
the Killing form for special linear algebras, and is half the
Killing form for special orthogonal algebras. It induces the
standard invariant symmetric bilinear form on the derived algebra
$\fg'$ as
\begin{equation}\label{blA}
\left(X\otimes\ld^k+\al\,c\mid
Y\otimes\ld^l+\beta\,c\right)=\dt_{k,-l}\frac{1}{r}\left(X\mid
Y\right)_0, \quad X, Y\in \mathcal{G}.
\end{equation}
On $\fg'$ the derivations $d_k$ reads
\begin{equation}\label{dkld}
d_k=-\frac{1}{r k_0}\ld^{1+r k_0 k}\frac{\od}{\od \ld}, \quad
k\in\Z,
\end{equation}
where $k_0=2$ whenever $\fg$ is of type $A_{2 n}^{(2)}$ and  $k_0=1$
otherwise.

\subsection{Tau functions of hierarchies of types $A_n^{(1)}$ and $D_n^{(1)}$}

\begin{exa} \label{exa-A1}
We realize the affine Kac-Moody algebra $\fg'$ of type $A_1^{(1)}$
by taking a set of Weyl generators as
\begin{align}
&e_0=\left(
              \begin{array}{cc}
                0 & \ld \\
                0 & 0 \\
              \end{array}
            \right)
, \quad f_0=\left(
              \begin{array}{cc}
                0 &  0 \\
                1/\ld & 0 \\
              \end{array}
            \right), \quad \al^\vee_0=\left(
              \begin{array}{cc}
                1 & 0 \\
                0 & -1 \\
              \end{array}
            \right)+c,  \label{weylA11}\\
&e_1=\left(
              \begin{array}{cc}
                0 & 0 \\
                1 & 0 \\
              \end{array}
            \right), \quad f_1=\left(
              \begin{array}{cc}
                0 & 0 \\
                1 & 0 \\
              \end{array}
            \right), \quad \al^\vee_1=\left(
              \begin{array}{cc}
                -1 & 0 \\
                0 & 1 \\
              \end{array}
            \right). \label{weylA12}
\end{align}
Let $\Ld=e_0+e_1$. The set of exponents is $E=\Z^{\mathrm{odd}}$,
and the principal Heisenberg subalgebra $\mathfrak{s}$ has a basis
\[
\{c, \Ld_j=\Ld^j\in\fg'^j\mid j\in \Z^{\mathrm{odd}}\}.
\]

The operator $\sL$ in \eqref{msL} is gauge equivalent to the
following canonical form
\begin{equation}
\sL^{\mathrm{can}}=D+\Ld+q^{\mathrm{can}}=D+\left(
        \begin{array}{cc}
          0 &  \ld \\
          1 & 0 \\
        \end{array}
      \right)+\left(
        \begin{array}{cc}
          0 &  -u \\
          0 & 0 \\
        \end{array}
      \right),
\end{equation}
let us compute the functions $U$ and $H$ determined as in
Proposition~\ref{thm-dr} by
\begin{align}\label{}
&D+\Ld+q^{\mathrm{can}}=e^{\ad_U}(D+\Ld+H), \label{con1} \\
&\big(e^{\ad_U}\Ld_j\big)_c=0, \quad j=1, 3, 5, \dots  \label{con2}.
\end{align}
First of all, we decompose $U$ and $H$ according to the principal
gradation $\fg'=\bigoplus_{k\in\Z}\fg'^k$ as
\begin{align}\label{}
U=&U_{-1}+U_{-2}+U_{-3}+U_{-4}+\cdots \nn\\
=& \left(
        \begin{array}{cc}
          0 &  a_1 \\
          b_1/\ld & 0 \\
        \end{array}
      \right)+\left(
        \begin{array}{cc}
          -a_2/\ld &  0 \\
          0 & a_2/\ld \\
        \end{array}
      \right)+\left(
        \begin{array}{cc}
          0 &  a_3/\ld \\
          b_3/\ld^2 & 0 \\
        \end{array}
      \right) \nn\\
      & +\left(
        \begin{array}{cc}
          -a_4/\ld^2 &  0 \\
          0 & a_4/\ld^2 \\
        \end{array}
      \right)+\cdots, \\
H=&-\frac{h_1}{2}\Ld_{-1}-\frac{h_3}{2}\Ld_{-3}-\cdots,
\end{align}
where $a_i$, $b_i$ and $h_i$ are scalar functions. Now substitute
them into \eqref{con1} and compare terms with equal principal
degrees. Clearly, the case of  degree $1$ is trivial. The first
nontrivial equation is
\begin{equation}\label{}
\mathrm{degree}\,0: \quad 0=[U_{-1}, \Ld]=(a_1-b_1)\,\diag(1,
-1)-b_1\cdot c.
\end{equation}
Hence $a_1=b_1=0$, i.e., $U_{-1}=0$. Note that the equality
\eqref{con2} with $j=1$ holds automatically. Secondly, we have
\begin{align}\label{}
\mathrm{degree}\,-1: \quad \left(
        \begin{array}{cc}
          0 &  -u \\
          0 & 0 \\
        \end{array}
      \right)=&[U_{-2},
\Ld]-\frac{h_1}{2}\Ld_{-1} \nn\\
=&\left(
        \begin{array}{cc}
          0 &  -2 a_2 \\
          2 a_2/\ld & 0 \\
        \end{array}
      \right)-\frac{h_1}{2}\left(
        \begin{array}{cc}
          0 &  1 \\
          1/\ld & 0 \\
        \end{array}
      \right),
\end{align}
which implies
\[
a_2=\frac{u}{4}, \quad h_1=u.
\]
Then it comes
\begin{align}
\mathrm{degree}\,-2: \quad
        0=&[U_{-3},
\Ld]-\pd_x U_{-2} \nn\\
=&(a_3-b_3)\,\diag\left(1/\ld,
-1/\ld\right)-\frac{u_x}{4}\diag\left(-1/\ld, 1/\ld\right).
\label{eqU3}
\end{align}
Here and below the subscript ``$x$'' stands for the partial
derivative with respective to it, for instance, $u_x=\pd_x u$ and
$u_{xx}=\pd_x^2 u$. The condition \eqref{con2} with $j=3$ reads
\begin{equation}\label{eqcon3}
0=[U_{-3}, \Ld_3]_c=-a_3-2\,b_3.
\end{equation}
Equations \eqref{eqU3} and \eqref{eqcon3} determine $a_3$ and $b_3$
uniquely:
\[
a_3=-\frac{u_x}{6}, \quad b_3=\frac{u_x}{12}.
\]
Subsequently,
\begin{align}
\mathrm{degree}\,-3: \quad
        0=&[U_{-4},
\Ld]+\frac{1}{2}[U_{-2},[U_{-2}, \Ld]] -\pd_x U_{-3}+ [U_{-2},
-\frac{h_1}{2}\Ld_{-1}]-\frac{h_3}{2}\Ld_{-3}
\end{align}
implies
\[
a_4=\frac{u^2}{8}+\frac{u_{xx}}{16}, \quad
h_3=\frac{u^2}{4}+\frac{u_{xx}}{12}.
\]

Lemma~\ref{thm-UN} shows that the Hamiltonian densities $h_j$ are
independent of the choice of gauge slice of $\sL$. By using
\eqref{DStau}, the tau function $\tau$ of the Drinfeld-Sokolov
hierarchy satisfies (note $\pd_{t_1}=\pd_x$)
\begin{align}\label{tauA11}
&\frac{\pd^2\log\tau}{\pd x^2}=\frac{1}{2}h_1=\frac{1}{2}u, \\
&\frac{\pd^2\log\tau}{\pd x\pd
t_3}=\frac{3}{2}h_3=\frac{3}{8}u^2+\frac{1}{8}u_{xx}.
\end{align}
They imply the KdV equation:
\begin{equation}\label{}
\frac{\pd u}{\pd t_3}=\frac{3}{2}u\,u_x+\frac{1}{4}u_{xxx}
\end{equation}

One can also fix the gauge slice of $\sL$ such that the dressing
operator $\Ta$ takes the form \eqref{Ta}. In fact,
\begin{equation}\label{}
\Ta_0=e^U e^\Om=e^{X_0+\ld^{-1}X_{1}+\ld^{-2}X_{2}+\dots},
\end{equation}
where $\ld^{-k}X_k$ lies in $\fg'_{-k}$ and particularly
\[
X_0=\left(
      \begin{array}{cc}
        0 & \om_1 \\
        0 & 0 \\
      \end{array}
    \right).
\]
We let $N=-X_0$, then $\Ta=e^N \Ta_0$ has the form \eqref{Ta}.
Noting $u=2\,\pd_x\om_1$, it is straightforward to calculate the
corresponding gauge slice:
\begin{equation}\label{}
\sL=e^{-\ad_{X_0}}\sL^{\mathrm{can}}=D+\Ld+\left(
      \begin{array}{cc}
        -\om & -\om^2-\om_x \\
        0 & \om \\
      \end{array}
    \right),
\end{equation}
in which we write $\om=\om_1$ to simply notations.
\end{exa}

\begin{exa}\label{exa-An}
Let $\fg'$ be an affine Lie algebra of type $A_n^{(1)}$. It can be
realized by choosing Weyl generators as follows \cite{DS, Kac}:
\begin{align}
&e_0=\ld\,e_{1,n+1}, \quad
 e_i=e_{i+1,i}~~~ (1\le i\le n), \label{eA} \\
&f_0=\frac{1}{\ld}e_{n+1,1},\quad f_i=e_{i,i+1}~~~ (1\le i\le n),\\
&\al^\vee_i=[ e_i,f_i] ~~~(0\le i\le n), \label{alvA}
\end{align}
where $e_{i,j}$ is the $(n+1)\times(n+1)$ matrix with its
$(i,j)$-component being $1$ and the others being zero. One has
$\Ld=e_0+e_1+\cdots+e_n$. A basis of the principal Heisenberg
subalgebra $\mathfrak{s}$ is $\{c, \Ld_j=\Ld^j\in\fg'^j\mid j\in
E\}$ with $E=\Z\setminus (n+1)\Z$ being the set of of exponents of
$\fg'$.

Take the operator $\sL$ in \eqref{sLgp} as the following canonical
form
\begin{equation}
\sL^{\mathrm{can}}=D+\Ld+q^{\mathrm{can}}, \quad
q^{\mathrm{can}}=\left(
        \begin{array}{ccccc}
          0 & \cdots & 0 & 0 & -u_n \\
           & \ddots & \vdots & \vdots & \vdots \\
           &  & 0 & 0 & -u_2 \\
           &  & & 0 & -u_1 \\
           &  & & & 0 \\
        \end{array}
      \right).
\end{equation}
According to \cite{DS}, the equations in \eqref{Lt2} can be
represented equivalently as
\begin{equation}\label{GD}
\frac{\pd L}{\pd t_j}=[(L^{j/(n+1)})_+,L],\quad j\in E_+,
\end{equation}
where
\begin{align}
&L=D^{n+1}+u_1 D^{n-1}+\dots+u_{n-1} D+u_n, \label{LAn0} \\
&L^{1/(n+1)}=D+v_1 D^{-1}+v_2 D^{-2}+\cdots, \nn
\end{align}
$(L^{j/(n+1)})_+$ means the differential part of the operator
$L^{j/(n+1)}$, and the multiplication of two pseudo-differential
operators is defined by
\[
u D^k\cdot v D^l=\sum_{m\ge0}u D^m(v) D^{k+l-m}.
\]
Clearly $\pd/\pd t_1=\pd/\pd x$. The hierarchy \eqref{GD} is just
the Gelfand-Dickey hierarchy \cite{GD76}, or the $n$th KdV
hierarchy.

It is known that the Gelfand-Dickey hierarchy \eqref{GD} carries two
compatible Hamiltonian structures, with \eqref{dspoi2} usually being
called the second one. The densities of the Hamiltonian functionals
are
\begin{equation}\label{GDh}
\hat{h}_j=\frac{n+1}{j}\res\,L^{j/(n+1)}, \quad j\in E_+.
\end{equation}
Note that the residue of a pseudo-differential operator is defined
by~$\res\sum_{i\in\Z}a_i\,D^i=a_{-1}$. It is easy to check
\begin{equation}\label{GDht}
\frac{j}{n+1}\frac{\pd \hat{h}_j}{\pd t_i}=\frac{i}{n+1}\frac{\pd
\hat{h}_i}{\pd t_j}, \quad i,j\in E_+.
\end{equation}
Given any solution of the hierarchy \eqref{GD}, there locally exits
a tau function $\hat{\tau}$ such that
\begin{equation}\label{GDtau}
\frac{\pd^2\log\hat{\tau}}{\pd t_i\,\pd
t_j}=\frac{j}{n+1}\pd_x^{-1}\frac{\pd\hat{h}_j}{\pd t_i}, \quad i,
j\in E_+.
\end{equation}
When $n=1$, this is just the case of the KdV hierarchy reviewed in
Section~1.


\begin{prp} \label{thm-tauA}
For the Drinfeld-Sokolov hierarchy of type $A_n^{(1)}$, the tau
function $\hat\tau$ in \eqref{GDtau} coincides\,\footnote{In the
present paper we say that two tau functions are the same if their
logarithms differ by addition of a linear function of the time
variables.} with $\tau$ defined in \eqref{DStau}.
\end{prp}
\begin{prf}
It is sufficient to identify the second-order derivatives of
$\log\hat{\tau}$ and $\log\tau$ with respect to the time variables.

On the one hand,
\begin{equation}\label{tauhxx}
\frac{\pd^2\log\hat{\tau}}{\pd
x^2}=\frac{1}{n+1}\hat{h}_1=\res\,L^{1/(n+1)}=\frac{u_1}{n+1}.
\end{equation}
On the other hand, for $\sL=\sL^{\mathrm{can}}$ we do the
calculation as in the proof of Proposition~\ref{thm-dr}. Equation
\eqref{Um1} now reads
\begin{equation*}\label{}
[U_{-1},\Ld]=q_0^{\mathrm{can}}=0,
\end{equation*}
which implies $U_{-1}=0$. The first equation in \eqref{HUk} is
\begin{equation*}\label{}
H_{-1}+[U_{-2},\Ld]=q_{-1}^{\mathrm{can}}=-u_1\,e_{n,n+1},
\end{equation*}
hence
\begin{align}\label{}
\frac{\pd^2\log\tau}{\pd x^2}=&\frac{(-\Ld_1\mid
H)}{(\Ld_1\mid\Ld_{-1})}
\nn\\
=&\frac{1}{(\Ld_1\mid\Ld_{-1})}(-\Ld_1\mid [\Ld,
U_{-2}]+q_{-1}^{\mathrm{can}})
 \nn\\
=&\frac{1}{(\Ld_1\mid\Ld_{-1})}(-\Ld_1\mid -u_1\,e_{n,n+1}) \nn\\
=&\frac{u_1}{n+1}. \label{tauxx}
\end{align}
It follows from \eqref{tauhxx} and \eqref{tauxx} that
\[
\frac{\pd}{\pd x}\left(\frac{\pd^2\log\tau}{\pd x\,\pd
t_j}-\frac{\pd^2\log\hat\tau}{\pd x\,\pd t_j}\right)=0, \quad j\in
E_+.
\]
The difference between the parentheses is a differential polynomial
in $u_1, \cdots, u_n$ without free term, thus it vanishes indeed. In
the same way, we derive
\begin{equation*}\label{}
\frac{\pd^2\log\tau}{\pd t_i\,\pd t_j}-\frac{\pd^2\log\hat\tau}{\pd
t_i\,\pd t_j}=0, \quad i,j\in E_+.
\end{equation*}
The proposition is proved.
\end{prf}

\end{exa}

\begin{exa}\label{exa-Dn}
Assume the affine Lie algebra $\fg'$ is of type $D_n^{(1)}$, and it
is realized as in \cite{LWZ}, see also \cite{DS, Wu}. We choose a
set of generators of the principal Heisenberg algebra $\mathfrak{s}$
as
\begin{align}\label{}
\Ld_k=\sqrt{2}\,\Ld^k&, \quad \Ld_{-k}=\sqrt{2}\,\Ld^{-k}, \\
\Ld_{k(n-1)'}=-\sqrt{2n-2}\,\Gm^k&, \quad
\Ld_{-k(n-1)'}=-\sqrt{2n-2}\,\Gm^{-k}
\end{align}
for $k\in\Zop$, where  $\Ld$ and $\Gm$ are two certain $2n\times 2n$
matrices given in \S\,4.2 of \cite{LWZ}. Note that the constant
$\nu=\sqrt{2}$, and that $k(n-1)$ are double exponents of $\fg'$
whenever $n$ is even.

Introduce a pseudo-differential operator
\begin{equation}\label{eq-L}
L=D^{2n-2}+\frac1{2}\sum_{i=1}^{n-1}D^{-1}\left(u_i
D^{2i-1}+D^{2i-1}u_i\right)+D^{-1}\rho D^{-1}\rho,
\end{equation}
There are uniquely two operators
\[
P=D+v_1 D^{-1}+v_2 D^{-2}+\cdots, \quad Q=D^{-1}\rho+\hat{v}_1
D+\hat{v}_2 D^2+\cdots
\]
such that $P^{2n-2}=L=Q^2$. The following integrable hierarchy are
well defined \cite{DS, LWZ}:
\begin{align}\label{DSD}
&\frac{\pd L}{\pd {t}_k}=\left[\sqrt{2}\,(P^k)_+, L\right], \quad
\frac{\pd L}{\pd t_{k(n-1)'}}=\left[-\sqrt{2n-2}\,(Q^k)_-, L\right],
\qquad k\in\Zop,
\end{align}
which is equivalent to the Drinfeld-Sokolov hierarchy \eqref{Lt2} of
type $D_n^{(1)}$ with $q^{\mathrm{can}}$ chosen in \cite{LWZ}.

The hierarchy \eqref{DSD} carries a bi-Hamiltonian structure, with
Hamiltonian densities being tau symmetric. Hence a tau function
$\hat{\tau}$ is defined by
\begin{equation}
\od\left(2\,\pd_x\log\hat{\tau}\right)=\sum_{k\in\Zop}
\left(\sqrt{2}\,\res\,P^k\,\od t_k+\sqrt{2n-2}\,\res\,Q^k\,\od
t_{k(n-1)'}\right). \label{Dtau}
\end{equation}
With the same method as for Proposition~\ref{thm-tauA}, we obtain
the following
\begin{prp} \label{thm-tauD}
For the Drinfeld-Sokolov hierarchy of type $D_n^{(1)}$, the tau
functions  $\hat\tau$ in \eqref{Dtau} and $\tau$ in \eqref{DStau}
coincide.
\end{prp}
\end{exa}
Based on this proposition, it is easy to see that the Virasoro
symmetries \eqref{Vktau} are consistent with those derived in
\cite{Wu-vir} with skills of pseudo-differential operators.

\begin{rmk}
In not so straightforward a way, Propositions~\ref{thm-tauA} and
\ref{thm-tauD} can be considered as corollaries of
Theorem~\ref{thm-KW}  by using relevant results in \cite{KW, LWZ}
that both $\tau$ and $\hat{\tau}$ are solutions of the corresponding
Kac-Wakimoto hierarchies.
\end{rmk}

\subsection{Obstacles in linearizing Virasoro symmetries}

\begin{exa} \label{exa-A1O}
We realize the affine Kac-Moody algebra $\fg'$ of type $A_1^{(1)}$
as in Example~\ref{exa-A1}. The simple Lie algebra
$\mathring{\fg}=\mathfrak{sl}_2$ contains the following elements of
$2\times2$ matrices:
\begin{align*}
&E_0=e_{1,2}, \quad F_0=e_{2,1}, \quad H_0=e_{1,1}-e_{2,2}, \\
&E_1=e_{2,1}, \quad F_1=e_{1,2}, \quad H_1=-e_{1,1}+e_{2,2},
\end{align*}
which correspond to the Weyl generators
\eqref{weylA11}--\eqref{weylA12}. The the derivations on $\fg'$ are
given by \eqref{dkld} with $r=k_0=1$, i.e.,
\[
d_k=-\ld^{k+1}\frac{\od}{\od \ld}, \quad k\in\Z.
\]
According to \eqref{Hs}, one has $H_{\rs^1}=\dfrac{1}{2}H_1$, and
then
\begin{equation}\label{}
H_{\rs^1}(k;\rs^0)=\dfrac{\ld^k}{2}H_1.
\end{equation}

Now we use the data in Example~\ref{exa-A1} to compute the obstacles
of Virasoro symmetries for the Drinfeld-Sokolov hierarchy associated
to $\fg'$. After a straightforward calculation, we have
\begin{align}\label{}
O_1'=&2\left(e^{\ad_U}d_1-d_1\right)_c
-\left(e^{\ad_U}H_{\rs^1}(1;\rs^0)\right)_c=0+\frac{1}{4}h_1
=\frac{1}{2}\frac{\pd^2\log\tau}{\pd t_1^2}, \\
O_2'=&2\left(e^{\ad_U}d_2-d_2\right)_c
-\left(e^{\ad_U}H_{\rs^1}(2;\rs^0)\right)_c=\frac{3}{2}h_3
=\frac{\pd^2\log\tau}{\pd t_1 \pd t_3}.
\end{align}
Hence $O_1=O_2=0$; furthermore, by using the Virasoro commutation
relation we obtain $O_k=0$ for all $k\ge3$. This fact agrees with
Corollary~\ref{thm-okADE}, which confirms directly the linearization
of Virasoro symmetries for the KdV hierarchy.
\end{exa}

\begin{exa} \label{exa-B2}
Let $\fg$ be the Kac-Moody algebra of type $B_2^{(1)}$ (isomorphic
to $C_2^{(1)}$), whose Cartan matrix is
\[
A=\left(
        \begin{array}{ccc}
          2 & 0 &  -1 \\
          0 & 2 & -1 \\
          -2 & -2 & 2 \\
        \end{array}
      \right).
\]
The elements $E_i$, $F_i$ and $H_i$ in subsection~4.1 can be
realized by $5\times5$ matrices as
\begin{align}\label{}
& E_0=\frac{1}{2}(e_{1,4}+e_{2,5}), \quad E_1=e_{2,1}+e_{5,4}, \quad
E_2=e_{3,2}+e_{4,3}, \\
& F_0=2(e_{4,1}+e_{5,2}), \quad F_1=e_{1,2}+e_{4,5}, \quad
F_2=2(e_{2,3}+e_{3,4}), \\
& H_i=[E_i, F_i] \quad (i=0,1, 2).
\end{align}
These elements give a set of Weyl generators according to
\eqref{weyls} for the homogeneous gradation $\rs^0$, hence $\fg(A;
\rs^0)$ is realized. The derivations in \eqref{dkld} are
\[
d_k=-\ld^{k+1}\frac{\od}{\od\ld}, \quad k\in\Z.
\]

Note $\Ld=E_1+E_2+\ld E_0$, and  $h=4$ is the Coxeter number. The
set of exponents of $\fg$ is $E=\Z^{\mathrm{odd}}$, and the
generators normalized by \eqref{dLd} of the principal Heisenberg
subalgebra $\mathfrak{s}$ are
\begin{equation}\label{}
\Ld_k=\sqrt{2}\,\Ld^k, ~~ \Ld_{-k}=\sqrt{2}\,(\ld^{-1}\Ld^3)^k,
\quad k\in\Zop.
\end{equation}
The operator $\sL$ \eqref{msL} is gauge equivalent to the canonical
form
\begin{equation}
\sL^{\mathrm{can}}=D+\Ld+q^{\mathrm{can}}, \quad
q^{\mathrm{can}}=-u\,(e_{1,2}+e_{4,5})-v\,(e_{1,4}+e_{2,5}).
\end{equation}
In the same way as before, we compute the functions
$U=U_{-1}+U_{-2}+\dots$ and $H$ given by Proposition~\ref{thm-dr}.
As a result,
\begin{align}\label{}
&U_{-1}=0, \\
&U_{-2}=\frac{3 u}{4}(-e_{1,3}+e_{3,5})+\frac{u}{2  \ld
}(-e_{3,1}+e_{5,3}), \\
&U_{-3}=-\frac{5 u_{x}}{6}(e_{1,4}+e_{2,5})+
\frac{u_{x}}{3\ld}(e_{2,1}+e_{5,4})-\frac{u_{x}}{6
\ld}(e_{3,2}+e_{4,3}), \\
&U_{-4}=\frac{u^2+4 \left(v+u_{xx}\right)}{4  \ld
}(-e_{1,1}+e_{5,5})+\frac{2 v+u_{xx}}{4  \ld }(-e_{2,2}+e_{4,4}); \\
 &H=-\frac{u}{2\sqrt{2}}\Ld_{-1}-\frac{3
u^2+12 v+10 u_{xx}}{24\sqrt{2}}\Ld_{-3}+\dots.
\end{align}
Note $\pd/\pd t_1=\sqrt{2}\,\pd/\pd x$, hence
\begin{align}\label{}
&\frac{\pd^2\log\tau}{\pd t_1^2}=\sqrt{2}\frac{(-\Ld_1\mid
H)}{(\Ld_1\mid\Ld_{-1})}=\frac{u}{2}, \\
&\frac{\pd^2\log\tau}{\pd t_1\pd t_3}=\sqrt{2}\frac{3(-\Ld_3\mid
H)}{(\Ld_3\mid\Ld_{-3})}=\frac{1}{8} \left(3 u^2+12 v+10
u_{xx}\right).
\end{align}

Clearly,
\begin{align}\label{}
&H_{\rs^1}=(H_1, H_2)\left(\mathring{A}^T\right)^{-1}(1, 1)^T =2
H_1+\frac{3}{2}H_2, \\
&H_{\rs^1}(k;\rs^0)=\ld^k\,H_{\rs^1}, \quad k\in\Z.
\end{align}
A straightforward calculation leads to
\begin{align}
O_1'=&4(e^{\ad_U}d_1-d_1)_c- \left(e^{\ad_U}H_{\rs^1}(1;\rs^0)\right)_c \nn\\
=&\frac{7 u^2}{8}+\frac{5 v}{2}+\frac{9 u_{xx}}{4} \nn\\
=&\frac{5}{3}\frac{\pd^2\log\tau}{\pd t_1\,\pd t_3}
+\frac{1}{6}\frac{\pd^4\log\tau}{\pd\, t_1^4}
+\left(\frac{\pd^2\log\tau}{\pd\, t_1^2}\right)^2.
\end{align}
It implies $O_1\ne0$ in \eqref{tauvir}, namely, the symmetry
$\pd/\pd\beta_1$ acting on the tau function cannot be linearized.
\end{exa}

\begin{exa}\label{exa-A2}
Based on the matrix realization of the simple Lie algebra $A_2$ in
Example~\ref{exa-An}, one chooses Weyl generators of the twisted
affine Lie algebra $\fg'$ of type $A_2^{(2)}$ as the following
$3\times3$ matrices:
\begin{align*}\label{}
&e_0=\ld(e_{2,1}+e_{3,2}), \quad e_1=e_{1,3}, \\
&f_0=\frac{2}{\ld}(e_{1,2}+e_{2,3}), \quad f_1=e_{3,1}, \\
&\al_0^\vee=H_0+c=-2\,e_{1,1}+2\,e_{3,3}+c, \\
&\al_1^\vee=H_1=e_{1,1}- e_{3,3}.
\end{align*}
The derivations \eqref{dkld} corresponding to the homogeneous
realization are
\[
d_k=-\frac{1}{4}\ld^{4k+1}\frac{\od}{\od\ld}, \quad k\in\Z.
\]

Let $\Ld=e_0+e_1$. The Coxeter number is $3$, and the principal
Heisenberg subalgebra $\mathfrak{s}$ contains a set of generators
\begin{equation}\label{}
\Ld_j=\sqrt{2}\,\Ld^j, \quad j\equiv\pm1 \mod 6,
\end{equation}
which satisfy the normalization condition \eqref{dLd}.

Take the canonical form of the operator $\sL$ as
\[
\sL^{\mathrm{can}}=D+\Ld+q^{\mathrm{can}}, \quad
q^{\mathrm{can}}=-u\,e_{3,1}.
\]
We compute the matrix-value functions $U$ and $H$ according to
Proposition~\ref{thm-dr}, and have
\begin{align}\label{}
U=&U_{-2}+U_{-3}+U_{-4}+U_{-5}+U_{-6}+\dots \nn\\
 =&\frac{u}{3\ld}(e_{2,1}-e_{3,2})+\frac{u_x}{9\ld^2}(-e_{1,1}+2e_{2,2}-e_{3,3})
 +\frac{u^2+2 u_{xx}}{18\ld^3}(-e_{1,2}+e_{2,3}) \nn\\
 &+\left(-\frac{5 u u_x+ 3u_{xxx}}{45\ld^4} e_{1,3}
 +\frac{5 u u_x+ 4u_{xxx}}{90\ld^3}(e_{2,1}+e_{3,2}) \right) \nn\\
 &+\frac{5 u^3+18 u_x^2+30 u u_{xx}+12
 u_{xxxx}}{324\ld^4}(-e_{1,1}+e_{3,3})+\dots, \\
H=&  -\frac{h_1}{3}\Ld_{-1}-\frac{h_5}{3}\Ld_{-5}+\dots \nn\\
&=-\frac{u}{3\sqrt{2}}\Ld_{-1}+\frac{5 u^3+15 u u_{xx}+3
u_{xxxx}}{405\sqrt{2}}\Ld_{-5}+\dots.
\end{align}
Hence by using $\pd/\pd t_1=\sqrt{2}\,\pd/\pd x$ we obtain
\begin{equation}\label{A22tau}
\frac{\pd^2\log\tau}{\pd t_1^2}=\frac{u}{3}, \quad
\frac{\pd^2\log\tau}{\pd t_1\pd t_5}=-\frac{1}{81} \left(5 u^3+15 u
u_{xx}+3 u_{xxxx}\right).
\end{equation}

On the other hand, one has
\begin{equation}\label{}
H_{\rs^1}(4 k;\rs^0)=\frac{\ld^{4 k}}{2}H_{1}, \quad k\in\Z,
\end{equation}
then
\begin{equation}\label{}
O_1'=6(e^{\ad_U}d_1-d_1)_c-(e^{\ad_U}H_{\rs^1}(4;\rs^0))_c=
\frac{5\sqrt{2}}{3}h_5= \frac{\pd^2\log\tau}{\pd t_1\pd t_5}.
\end{equation}
Thus we obtain $O_1=0$, which agrees with Corollary~\ref{thm-okADE}.

\begin{rmk}
As an application of \eqref{A22tau}, one has the first nontrivial
equation of the $A_2^{(2)}$-hierarchy:
\begin{equation}\label{SK}
\frac{\pd u}{\pd t_5}=-\frac{1}{108}\frac{\pd}{\pd t_1}\left(20
u^3+30 u \frac{\pd^2 u}{\pd t_1^2}+3 \frac{\pd^4 u}{\pd
t_1^4}\right),
\end{equation}
which is also known as the Sawada-Kotera equation \cite{SK}. This
equation carries a generalized bi-Hamiltonian structure consisting
of a local and a nonlocal operators, as was shown by Fuchssteiner
and Oevel \cite{FO}. In fact, by using a perturbation approach we
proved that equation \eqref{SK} possesses exactly one local
Hamiltonian structure \cite{LWZ-ham} that coincides with the first
Hamiltonian structure in \cite{FO}.
\end{rmk}

\end{exa}

\section{Conclusion and remark}

We have obtained a unified characterization of tau function and
Virasoro symmetries for Drinfeld-Sokolov hierarchy associated to any
affine Kac-Moody algebra and the zeroth vertex of its Dynkin
diagram. This together with the results in \cite{DS, DLZ} suggests
us, although a complete proof is still missing, to divide these
Drinfeld-Sokolov hierarchies into three classes as in Table~1.
\begin{table}[h]
\caption{}\label{}
\begin{center}
{\setlength{\doublerulesep}{0.5pt}
\begin{tabular}{cccc}
\hline\hline
Class & Affine algebra $X_N^{(r)}$ & Hamiltonian structure & Virasoro symmetries  \\
\hline
\\
I & $A_n^{(1)}$, $D_n^{(1)}$, $E_{6, 7, 8}^{(1)}$  & bi-Hamiltonian
& linearizable
\\
\\
II & $B_n^{(1)}$, $C_n^{(1)}$, $F_4^{(1)}$, $G_2^{(1)}$   &
bi-Hamiltonian & \emph{non-linearizable}
\\
\\
III & \parbox[c]{0.21\textwidth}{ $A_{2n}^{(2)}$, $A_{2n-1}^{(2)}$,
$D_{n+1}^{(2)}$, $E_6^{(2)}$, $D_4^{(3)}$ } & \emph{Hamiltonian} &
linearizable
\\
\\
\hline\hline
\end{tabular}
}
\end{center}
\end{table}

The hierarchies in Class~I are bi-Hamiltonian and have linearized
Virasoro symmetries acting on the tau function. They coincide, up to
a rescaling of the time variables, with the topological hierarchies
constructed by Dubrovin and Zhang \cite{DZ, LRZ} associated to
semisimple Frobenius manifolds for ADE-type Weyl groups. Their tau
functions defined in \eqref{DStau} that admit the Virasoro
constraints give partition functions of 2\,D topological minimal
models, as well as Givental's total descendant potentials for simple
singularities, see \cite{DVV, Gi-A, GM, FGM, Wu} and references
therein.

Each hierarchy in Class~II also carries a bi-Hamiltonian structure
whose leading term is associated to a semisimple Frobenius manifold.
However, its Virasoro symmetries are expected to be
non-linearizable. In fact, the non-linearizability of Virasoro
symmetries for the $C_n^{(1)}$-hierarchies has been verified by us
\cite{CW} based on a generating function for $O_k$; for the
$B_n^{(1)}$-hierarchies it can be implied by the failure in reducing
the additional symmetries of the BKP hierarchy to symmetries of
$B_n^{(1)}$-hierarchies with the method of \cite{Wu-vir}, which and
the exceptional cases will be considered elsewhere. The meaning of
the obstacles in linearizing Virasoro symmetries is still far from
being well understood, which probably illustrates the observation of
the absence of a consistent higher-genus expansion beyond genus one
in topological minimal models associated to Lie algebras other than
ADE type \cite{EYY}.

From the viewpoint of tau functions and Virasoro symmetries, the
hierarchies in Class~III look very similar with those in Class~I.
The main difference between them is that, every hierarchy in
Class~III, such like the $A_2^{(2)}$-hierarchy, may not possess more
than one (local) Hamiltonian structure. It is an interesting
question whether there is some topological meaning or axiomatic
construction for them such like in \cite{DZ}.

As a byproduct, we have shown that the tau function of a hierarchy
in Class~I or III is a solution of the corresponding Kac-Wakimoto
hierarchy of bilinear equations. This is consistent with the
corresponding results for the case of types $A_n^{(1)}$ and
$D_n^{(1)}$ in the literature.

Finally, note that we have concentrated ourselves to the original
Drinfeld-Sokolov hierarchies, which is widely applied in some other
research areas. Since there are varies of generalizations of the
Drinfeld-Sokolov hierarchies, see, for example, \cite{FHM, FM,
dGHM}, for them it is natural to ask whether there is an analogous
characterization of tau function and Virasoro symmetries. By now a
positive answer can be seen for the so-called generalized
hierarchies of type~I in \cite{dGHM}, in which the principle
Heisenberg subalgebra is replaced by an arbitrary Heisenberg
subalgebra (see also \cite{CaW}). However, it is still unclear for
the other cases. We will study it on other occasions.

\vskip 0.5truecm \noindent{\bf Acknowledgments.} The author thanks
Professor  Youjin Zhang for his advising, and thanks Professors
Boris Dubrovin and Si-Qi Liu for many hours of discussions.  The
work was partially done when the author was a Marie Curie fellow of
the Istituto Nazionale di Alta Matematica hosted in SISSA, Trieste,
Italy, and partially supported by the ``Young SISSA Scientists'
Research Projects'' scheme 2012--2013.

\begin{appendices}
\section*{Appendix}
\section{Tau functions of modified Drinfeld-Sokolov hierarchies}

Let us consider the modified Drinfeld-Sokolov hierarchy associated
to an affine Lie algebra $\fg'=\bar\fg\oplus\C\,c$, which is defined
by \eqref{Lt2} with $\sL$ given by a function $q$ taking value in
the Cartan subalgebra of $\mathring{\fg}\subset\bar\fg$. This
particular gauge slice of $\sL$ is related to the others by gauge
transformations \eqref{gauge} that do not change the definition of
$\tau$ in \eqref{DStau}, hence $\tau$ also serves as a tau function
of the modified Drinfeld-Sokolov hierarchy. We are to compare this
$\tau$ with those tau functions introduced by Enriquez and Frenkel
\cite{EF} and by Miramontes \cite{Mi}.

\subsection{Enriquez and Frenkel's tau function}

For the modified Drinfeld-Sokolov hierarchy associated to an
untwisted affine Lie algebra $\bar\fg$, Enriquez and Frenkel
\cite{EF} found the following Hamiltonian densities:
\begin{equation}\label{thj}
\tilde{h}_j=(\Ld_1\mid e^{\ad_U}\Ld_j), \quad j\in E_+,
\end{equation}
where $U$ is given in Proposition~\ref{thm-dr0}. These densities
satisfy
\[
 \frac{\pd\tilde{h}_j}{\pd t_i}=\frac{\pd\tilde{h}_i}{\pd t_j}=\frac{\pd
H_{i,j}}{\pd x}
\]
with $H_{i,j}$ being differential polynomials in $q$; in other
words, they are tau-symmetric. There locally exists a tau function
$\tilde{\tau}$ such that
\begin{equation}\label{tauEF}
\frac1{(\Ld_1\mid\Ld_{-1})}\tilde{h}_j=\frac{\pd^2\log\tilde{\tau}}{\pd
t_1\,\pd t_j}, \quad j\in E_+.
\end{equation}
In comparison with those equations in \S\,5.5 of \cite{EF}, here on
the left hand side of \eqref{tauEF} we assign a coefficient
$1/{(\Ld_1\mid\Ld_{-1})}$ to make $\tilde{\tau}$ independent of the
choice of the invariant symmetric bilinear form.

According to Proposition~\ref{thm-dr0}, the freedom of the function
$U$ does not change the densities $\tilde{h}_j$. Without lose of
generality, let us consider the derived algebra
$\fg'=\bar{\fg}\oplus\C\,c$ instead of $\bar\fg$, and fix $U$ as in
Proposition~\ref{thm-dr}.

\begin{prp}
For the modified Drinfeld-Sokolov hierarchy associated to any affine
Lie algebra $\fg'$, the tau functions defined by \eqref{tauEF} and
by \eqref{DStau} satisfy
\begin{equation}\label{tautaut}
\log\tilde{\tau}-\log\tau=\frac1{(\Ld_1\mid\Ld_{-1})}\left(\frac{\pd}{\pd
t_1}\right)^{-1}\big(\Ld_1\mid U_{-1}\big),
\end{equation}
where $U_{-1}\in C^\infty(\R,\fg'^{-1})$ is uniquely determined by
$[U_{-1},\Ld]=q$. Note that $(\Ld_1\mid U_{-1})$ is a nontrivial
linear combination of components of $q$.
\end{prp}
\begin{prf}
According to the principal gradation \eqref{prgr} on $\fg'$, we
write $U=U_{-1}+U_{-2}+\cdots$ with $U_{k}\in\fg'^k$. Since $q$
takes value in the Cartan subalgebra, then $[U_{-1},\Ld]=q$.

Recall
\begin{equation*}\label{eq1}
\sL_1=e^{\ad_U}\left(\frac{\pd}{\pd t_1}+\Ld_1-\sum_{j\in
E_+}\frac{1}{j}\frac{\pd\om_j}{\pd t_1}\Ld_{-j}-\om_1\cdot c\right).
\end{equation*}
In particular, the projection of $\sL_1-\pd/\pd t_1$ onto
$\fg'^{-1}$ vanishes, namely,
\begin{equation*}\label{}
-\frac{\pd U_{-1}}{\pd
t_1}+\left(e^{\ad_U}\Ld_1\right)_{-1}-\frac{\pd\om_1}{\pd
t_1}\Ld_{-1}=0.
\end{equation*}
Since $\om_1=\pd\log\tau/\pd t_1$, then
\begin{equation}\label{eq3}
(\Ld_1\mid e^{\ad_U}\Ld_1)=\frac{\pd}{\pd t_1}(\Ld_1\mid
U_{-1})+(\Ld_1\mid\Ld_{-1})\frac{\pd^2\log\tau}{\pd t_1^2}.
\end{equation}
Substitute \eqref{thj} and \eqref{tauEF} into the left hand side of
\eqref{eq3}, then the equality \eqref{tautaut} follows from
integrations. The proposition is proved.
\end{prf}

\begin{exa}\label{exa-EF}
Let $\fg'$ be the affine Lie algebra of type $A_1^{(1)}$. We realize
$\fg'$ as in Example~\ref{exa-A1}, and take
\begin{equation}\label{}
\sL=D+\Ld+q, \quad q=\diag(-v,v).
\end{equation}
It is straightforward to compute the functions $U$ and $H$ in
Proposition~\ref{thm-dr}. Then it follows from \eqref{tauEF} and
\eqref{DStau} that
\begin{align}\label{tauA1EF}
&\frac{\pd^2\log\tilde{\tau}}{\pd x^2}=\frac1{2}\tilde{h}_1=-\frac1{2}v^2, \\
&\frac{\pd^2\log\tau}{\pd
x^2}=\frac1{2}h_1=-\frac1{2}(v^2-v_x),\label{tauA1}
\end{align}
where we have used $\pd/\pd t_1=\pd/\pd x$. These two tau functions
satisfy
\begin{equation}\label{}
v=2\,\pd_x\log\frac{\tau}{\tilde{\tau}},
\end{equation}
which is an equivalent version of \eqref{tautaut}.

In fact, by solving the equation of gauge transformation
\begin{align}\label{}
&\left(
  \begin{array}{cc}
    1 & g \\
    0 & 1 \\
  \end{array}
\right)\left(D+\left(
                 \begin{array}{cc}
                   0 & \ld \\
                   1 & 0\\
                 \end{array}
               \right)+\left(
                         \begin{array}{cc}
                           -v & 0 \\
                           0 & v \\
                         \end{array}
                       \right) \right)\left(
  \begin{array}{cc}
    1 & -g \\
    0 & 1 \\
  \end{array}
\right) \\
=&D+\left(
                 \begin{array}{cc}
                   0 & \ld \\
                   1 & 0\\
                 \end{array}
               \right)+\left(
                         \begin{array}{cc}
                           0 & -u \\
                           0 & 0 \\
                         \end{array}
                       \right),
\end{align}
one has $g=v$ and
\begin{equation}\label{}
u=-v^2+v_x.
\end{equation}
This is the well-known Miura transformation that relates the KdV and
the modified KdV equations. From \eqref{tauA1} we derive again
$u=2\pd_x^2\log\tau$, see \eqref{tauA11}.
\end{exa}

Enriquez and Frenkel's construction of tau function $\tilde\tau$ can
be formally extended to the modified Drinfeld-Sokolov hierarchies
associated to twisted affine algebras by \eqref{tautaut}. In fact,
the different tau functions $\tau$ and $\tilde{\tau}$ can be
interpreted by a general result in \cite{Mi}, see
Proposition~\ref{thm-taum} below.

\subsection{Miramontes' tau function}

Tau functions of the (generalized) modified Drinfeld-Sokolov
hierarchy were also constructed by Miramontes \cite{Mi}. His method
is based on a highest weight representation of the Kac-Moody group,
and the resulting tau function is related to a family of
conservation laws for the hierarchy.

Let us recall some statements in \cite{Mi}, following closely the
original notations used there. Given a gradation $\rs'=(s_0', s_1',
\dots, s_n')\in\Gm$ on the affine Lie algebra $\fg'$ (recall
\eqref{sgr}):
\begin{equation}\label{}
\fg'=\bigoplus_{j\in\Z}\fg'_{j\,[\rs']}.
\end{equation}
Similarly as before, we use notations like $\fg'_{\le
k\,[\rs']}=\sum_{j\le k}\fg'_{j\,[\rs']}$, and let the subscript
``$\le k\,[\rs']$'' stand for the projection  $\fg'\to\fg'_{\le
k\,[\rs']}$. Assume that an element $\Ld\in\fg'_{k\,[\rs']}$ ($k>0$)
is fixed, and it induces the following decomposition of subspaces:
\begin{equation}\label{}
\fg'=\kn\,\ad_\Ld+\im\,\ad_\Ld, \quad
\kn\,\ad_\Ld\cap\im\,\ad_\Ld=\C\,c.
\end{equation}
Here $\kn\,\ad_\Ld$ is a subalgebra consisting of elements that
commute with $\Ld$ modulo $\C\,c$; this subalgebra is generated by
$\Ld_j\in\fg'_{j\,[\rs']}$ ( $j\in E_\Ld\subset\Z$ ) such that
$[\Ld_j,\Ld_{-j}]\in\C\,c$.

Choose another gradation $\rs=(s_0,s_1,\dots,s_n)\preceq\rs'$ of
$\fg'$, namely, $s_i\le s_i'$ for $i=0,1,\dots,n$. With the
principal and homogeneous gradations on $\fg'$ replaced by the
gradations $\rs'$ and $\rs$ respectively, de Groot, Hollowood and
Miramontes \cite{dGHM} defined the generalized Drinfeld-Sokolov
hierarchies similarly as in Definition~\ref{def-DS}.

In more details, let
\begin{equation}\label{}
\mL=D+\Ld+q+w\cdot c, \quad q\in
C^{\infty}\left(\R,\fg'_{0\,[\rs]}\cap\fg'_{\le0\,[\rs']}\right),
\end{equation}
where the central part of $q$ is zero, and $w$ is an arbitrary
smooth function. Similar to Proposition~\ref{thm-dr0}, there exists
a function $Y\in C^{\infty}\left(\R,\fg'_{<0\,[\rs']}\right)$ such
that
\begin{equation}\label{LMi}
\mL=e^{\ad_Y}(D+\Ld+h^{\Phi})=\Phi(D+\Ld+h^{\Phi})\Phi^{-1}
\end{equation}
with $\Phi=e^Y$ and $h^{\Phi}\in
C^{\infty}\left(\R,\kn\,\ad_\Ld\cap\fg'_{\le0\,[\rs']}\right)$.
Moreover, the functions $Y$ and $h^\Phi$ are supposed to be
specified by the constraints (see (2.25) in \cite{Mi}):
\begin{equation}\label{conmi}
Y\in \fg'_{<0\,[\rs]}, \quad (h^{\Phi})_{\ge0\,[\rs]}\in\C\,c.
\end{equation}

Introduce operators
\begin{equation}\label{LLj}
\mL_j=\frac{\pd}{\pd t_j}+A_j, \quad
A_j=(\Phi\Ld_j\Phi^{-1})_{\ge0\,[\rs]}+w_j\cdot c, \quad j\in
E_{\Ld\,+},
\end{equation}
where $w_j$ are some priori functions. The flows of the generalized
Drinfeld-Sokolov hierarchy are defined by the following
zero-curvature equations
\begin{equation}\label{dsmi}
[\mL,\mL_j]=0, \quad j\in (E_{\Ld})_{\ge0}.
\end{equation}
These equations imply
\begin{equation}\label{Lh}
\Phi^{-1}\mL_j\Phi=\frac{\pd}{\pd t_j}+\Ld_j+h(j)+w_j\cdot c,
\end{equation}
where
\begin{equation*}\label{}
h(j)=\Phi^{-1}\frac{\pd\Phi}{\pd
t_j}-\Phi^{-1}(\Phi\Ld_j\Phi^{-1})_{<0\,[\rs]}\Phi \in
C^\infty\left(\R,\kn\,\ad_\Ld\cap\fg'_{<0\,[\rs']}\right).
\end{equation*}
In \cite{Mi} the hierarchy \eqref{dsmi} is restricted by $t_1=\nu x$
and $\mL_1=\nu\mL$ with some normalization constant $\nu$. Hence one
has  $\nu\,h^\Phi=h(1)+w_1\cdot c$.

For every auxiliary gradation
$\mathrm{m}=(m_0,m_1,\dots,m_n)\preceq\rs$, there exists a
derivation $d_0^{(\mathrm{m})}$ in the Kac-Moody algebra $\fg$ whose
derived algebra is $\fg'$. For the sake of simplifying notations, we
redenote $d_0^{(\mathrm{m})}$ as $d_{\mathrm{m}}$; it satisfies (see
Section~\ref{sec-vir})
\begin{equation}\label{}
[d_{\mathrm{m}}, e_i]=m_i\,e_i, \quad [d_{\mathrm{m}},
f_i]=-m_i\,f_i, \quad [d_{\mathrm{m}}, \al^\vee_i]=0
\end{equation}
for $i=0,1,\dots,n$. The standard bilinear form on $\fg'$ is
extended to $\fg$ by
\begin{equation}\label{}
(d_{\mathrm{m}}\mid d_{\mathrm{m}})=0, \quad
(d_{\mathrm{m}}\mid\al^\vee_i)=\frac{k_i}{k_i^\vee}m_i ~~\hbox{ for
}~~ i=0,1,\dots,n.
\end{equation}
Note that $ (d_{\mathrm{m}}\mid c)=N_{\mathrm{m}}$ with
$N_{\mathrm{m}}=\sum_{i=0}^{n}k_i\,m_i$.

One can define the following conserved densities of the hierarchy
\eqref{dsmi} (see equation~(3.8) in \cite{Mi}):
\begin{equation}\label{Jjk}
J_{j,k}[\mathrm{m}]=\frac{N_{\rs'}}{k\,N_{\mathrm{m}}}\big(d_{\mathrm{m}}\mid
[\Phi\Ld_k\Phi^{-1},A_j]\big), \quad j,k\in (E_{\Ld})_{\ge0}.
\end{equation}
These conserved densities are related to a tau function
$\tau_{\bar{\mathrm{m}}}$ as (cf. equation~(4.13) in \cite{Mi})
\begin{equation}\label{Jtau}
J_{j,k}[\mathrm{m}]=-\frac{N_{\rs'}}{k}\frac{\pd^2\log\tau_{\bar{\mathrm{m}}}}{\pd
t_j\,\pd t_k}, \quad j, k\in (E_{\Ld})_{\ge0}.
\end{equation}
Here $\bar{\mathrm{m}}=(m_0 k_0/k_0^\vee, \dots, m_n k_n/k_n^\vee)$,
it induces an integrable highest weight representation
$L(\bar{\mathrm{m}})$ of $\fg$ with highest weight vector
$|v_{\bar{\mathrm{m}}}\ra$, and $\tau_{\bar{\mathrm{m}}}$ is defined
by an element of the orbit of $|v_{\bar{\mathrm{m}}}\ra$ acted by
the Kac-Moody group, see \S~4.1 and the appendix in \cite{Mi} for
details.

Now let us compare the tau function in \eqref{Jtau} with those in
\eqref{tautaut} for the modified Drinfeld-Sokolov hierarchies.
Henceforth we fix both $\rs$ and $\rs'$ to be the principal
gradation $\rs^1$. In this case $E_\Ld=E$ is the set of exponents of
$\fg'$, and $\kn\,\ad_\Ld$ is the principal Heisenberg subalgebra
$\mathfrak{s}$. Choose generators $\Ld_j$ normalized by \eqref{dLd},
with $\Ld_1=\nu\Ld=\nu\sum_{i=0}^n e_i$. Let
\[
\mL_j=\sL_j ~~\hbox{ for }~~ j\in E_+, \quad Y=U, \quad h(1)=\nu H
\]
with $U$ and $H$ determined as in Proposition~\ref{thm-dr}. Observe
that both $Y$ and $h^\Phi$ admit the constraint \eqref{conmi}.

\begin{prp}\label{thm-taum}
For the modified Drinfeld-Sokolov hierarchy associated to affine Lie
algebra $\fg'$ with $\rs=\rs'=\rs^1$ being the principal gradation,
the two tau functions in \eqref{tautaut} can be considered as
particular cases of tau functions in \eqref{Jtau}. More precisely,
the following equalities hold true
\begin{equation}\label{taum}
\log\tau_{\bar{\mathrm{m}}}=\left\{\begin{array}{cc}
                              \log\tau,  & \mathrm{m}=(1,0,\dots,0);
                              \\ \\
                              \log\tilde{\tau}, &
                              \mathrm{m}=(1,1,\dots,1).
                            \end{array}\right.
\end{equation}
\end{prp}
\begin{prf}
Since $N_{\rs^1}$ is equal to the Coxeter number $h$, one recasts
\eqref{Jjk} to
\begin{align}\label{Jjk2}
J_{j,k}[\mathrm{m}]=&\frac{h}{k\,N_{\mathrm{m}}}\left(d_{\mathrm{m}}\mid\bigg[\Phi\Ld_k\Phi^{-1},\mL_j-\frac{\pd}{\pd
t_j}\bigg]\right) \nn\\
=&\frac{h}{k\,N_{\mathrm{m}}}\left(d_{\mathrm{m}}\mid\Phi\bigg[\Ld_k,
\frac{\pd}{\pd t_j}+\Ld_j+h(j)+w_j\cdot c\bigg]\Phi^{-1}\right)
\nn\\
&+\frac{h}{k\,N_{\mathrm{m}}}\left(d_{\mathrm{m}}\mid\frac{\pd}{\pd
t_j}(\Phi\Ld_k\Phi^{-1})\right)\nn\\
=&\frac{h}{k}[\Ld_k,
h(j)]_c+\frac{h}{k\,N_{\mathrm{m}}}\frac{\pd}{\pd
t_j}\left(d_{\mathrm{m}}\mid\Phi\Ld_k\Phi^{-1}\right).
\end{align}

If the auxiliary gradation  $\mathrm{m}=(1,0,\dots,0)$ is the
homogeneous one, then $N_{\mathrm{m}}=k_0$. By virtue of
\begin{equation*}\label{}
(d_{\mathrm{m}}\mid\al^\vee_i)=\dt_{i0}\frac{k_i}{k_i^\vee}, \quad
(\Phi\Ld_k\Phi^{-1})_c=(e^{\ad_U}\Ld_k)_c=0,
\end{equation*}
the second term in \eqref{Jjk2} vanishes. Hence we have
\begin{equation}\label{}
\frac{k}{h}J_{1,k}[\mathrm{m}]=[\Ld_k,h(1)]_c=\frac{k}{(\Ld_k\mid
\Ld_{-k})}(\Ld_k\mid\nu H),
\end{equation}
that is, thanks to \eqref{Jtau} and \eqref{omhj},
\begin{equation}\label{taumtau}
-\frac{\pd^2\log\tau_{\bar{\mathrm{m}}} }{\pd t_1\,\pd
t_k}=-\frac{\pd^2\log\tau}{\pd t_1\,\pd t_k}.
\end{equation}

If $\mathrm{m}=(1,1,\dots,1)$ is the principal gradation, then
$N_{\mathrm{m}}=h$. By using \eqref{Jjk2} and
$(\Ld_j\mid\Ld_k)=\dt_{j,-k}\,h$ one has
\begin{align}\label{}
\frac{1}{h}J_{1,1}[\mathrm{m}]=&[\Ld_1,h(1)]_c+\frac1{h}\frac{\pd}{\pd
t_1}(d_{\mathrm{m}}\mid\Phi\Ld_1\Phi^{-1}) \nn\\
=&\frac{(\Ld_1\mid \nu
H)}{(\Ld_1\mid\Ld_{-1})}+\frac1{h}\frac{\pd}{\pd
t_1}(d_{\mathrm{m}}\mid[U_{-1},\Ld_1])
\nn\\
=&-\frac{\pd^2\log\tau}{\pd t_1^2}- \frac1{h}\frac{\pd}{\pd
t_1}([d_{\mathrm{m}},\Ld_1]\mid U_{-1})\nn\\
=&-\frac{\pd^2\log\tau}{\pd t_1^2}-
\frac{1}{(\Ld_1\mid\Ld_{-1})}\frac{\pd}{\pd t_1}(\Ld_1\mid U_{-1}).
\label{J11}
\end{align}
This together with \eqref{Jtau} and \eqref{tautaut} leads to
\begin{equation}\label{taumtaut}
\frac{\pd^2\log\tau_{\bar{\mathrm{m}}} }{\pd
t_1^2}=\frac{\pd^2\log\tilde{\tau}}{\pd t_1^2}.
\end{equation}

Therefore, the proposition follows from \eqref{taumtau} and
\eqref{taumtaut} by integration in the same way as to show
Proposition~\ref{thm-tauA}.
\end{prf}

The proof of this theorem also implies that, whenever $\rs$ is the
principal gradation, one can choose $n+1$ linearly independent
auxiliary gradations $\mathrm{m}\preceq\rs$, which correspond to
$n+1$ distinct tau functions $\tau_{\bar{\mathrm{m}}}$ of the
modified Drinfeld-Sokolov hierarchy. In particular, suppose $\fg'$
is of type $A^{(1)}_{1}$, then such two tau functions of the
modified KdV hierarchy are given in Example~\ref{exa-EF}.


\end{appendices}

\end{document}